\documentclass[sigconf]{acmart}

\usepackage{comment}
\usepackage{graphicx}
\usepackage{xcolor}
\usepackage{multirow}
\usepackage{siunitx}
\usepackage{todonotes}
\usepackage{hyperref}
\usepackage{amsmath}
\usepackage{amsfonts}
\usepackage{algorithmic}
\usepackage{subfig}
\usepackage{makecell}
\usepackage{svg}
\usepackage{tablefootnote}
\usepackage[flushleft]{threeparttable}
\usepackage{fixltx2e}
\usepackage{caption} 
\usepackage{soul}
\usepackage{nicefrac}

\usepackage{url}

\begin{document}



\title{Striking the Balance: GEMM Performance Optimization Across Generations of Ryzen™ AI NPUs}



\author{Endri Taka}
\authornote{Work performed during an internship at AMD, Inc.}
\affiliation{%
  \institution{The University of Texas at Austin}
  \city{Austin}
  \state{TX}
  \country{United States}
}
  \email{endri.taka@utexas.edu}

\author{Andre Roesti}
\affiliation{%
  \institution{Advanced Micro Devices, Inc.}
  \city{Longmont}
  \state{CO}
  \country{United States}
}
  \email{andre.roesti@amd.com}

\author{Joseph Melber}
\affiliation{%
  \institution{Advanced Micro Devices, Inc.}
  \city{Longmont}
  \state{CO}
  \country{United States}
}
  \email{joseph.melber@amd.com}

\author{Pranathi Vasireddy}
\affiliation{%
  \institution{Advanced Micro Devices, Inc.}
  \city{Longmont}
  \state{CO}
  \country{United States}
}
  \email{pvasired@amd.com}

\author{Kristof Denolf}
\affiliation{%
  \institution{Advanced Micro Devices, Inc.}
  \city{Longmont}
  \state{CO}
  \country{United States}
}
  \email{kristof.denolf@amd.com}

\author{Diana Marculescu}
\affiliation{%
  \institution{The University of Texas at Austin}
  \city{Austin}
  \state{TX}
  \country{United States}
  }
  \email{dianam@utexas.edu}



\renewcommand{\shortauthors}{Endri Taka et al.}

\begin{abstract}

The high computational and memory demands of modern deep learning (DL) workloads have led to the development of specialized hardware devices from cloud to edge, such as AMD's Ryzen™ AI XDNA™ NPUs.
Optimizing general matrix multiplication (GEMM) algorithms for these architectures is critical for improving DL workload performance.
To this end, this paper presents a \textit{common} systematic methodology to optimize GEMM workloads across the two current NPU generations, namely XDNA and XDNA2.
Our implementations exploit the unique architectural features of AMD's NPUs and address key performance bottlenecks at the system level.
End-to-end performance evaluation across various GEMM sizes demonstrates state-of-the-art throughput of up to 6.76 TOPS (XDNA) and 38.05 TOPS (XDNA2) for 8-bit integer (int8) precision.
Similarly, for brain floating-point (bf16) precision, our GEMM implementations attain up to 3.14 TOPS (XDNA) and 14.71 TOPS (XDNA2).
This work provides significant insights into key performance aspects of optimizing GEMM workloads on Ryzen AI NPUs.




\end{abstract}


\keywords{Computer Architecture, Deep Learning, Hardware Acceleration, Matrix Multiplication, Neural Processing Unit, Ryzen™ AI}

\maketitle

\section{Introduction}
\label{sec:Introduction}

The widespread adoption of deep learning (DL) applications, combined with their intensive computational requirements, has driven the emergence of specialized hardware platforms.
While cloud data centers continue to provide large-scale acceleration \cite{NVIDIA_H100, AMD_CDNA_4, TPUv42021, MAIA_microsoft_2024, Meta_second_gen_chip, Amazon_Inferentia2}, the increasing demand for energy-efficiency, low-latency, as well as enhanced privacy and security 
has motivated the integration of DL accelerators on the edge \cite{NPU_arch_MICRO_2024, Versal_AI_Edge_gen_2, Intel_NPU, Qualcomm_Hexagon_NPU, Nvidia_Jetson_nano}.
To this end, AMD released the Ryzen™ AI processors \cite{NPU_arch_MICRO_2024}, featuring multi-core CPUs, an integrated GPU, and a new neural processing unit (NPU).
The NPU features the XDNA™ architecture and serves as a dedicated DL accelerator integrated with x86 processors. 
The AMD NPU is an evolution of the AI Engine (AIE) architecture, utilized in Versal™ adaptive system-on-chip (SoC) platforms \cite{Versal_Hot_chips_2019, Versal_Hot_chips_2021} and Alveo™ V70 accelerator cards \cite{AMD_CES_2023}.
The NPU architecture leverages the characteristics of modern DL workloads, where compilers can determine most of the control flow statically.
In particular, AMD NPUs provide an explicit data movement architecture, which both reduces hardware complexity and enables high performance, while offering substantially higher energy-efficiency compared to dynamically scheduled architectures, such as CPUs and GPUs \cite{NPU_arch_MICRO_2024}.

As modern DL workloads are dominated by general matrix multiplication (GEMM) operations, several prior works have aimed to design and optimize GEMM workloads on the Versal platforms
\cite{MaxEVA_2023, CHARM_FPGA_2023, AutoMM_DAC_2023, Versal_vs_Stratix_FCCM_2024, Charm_v2_2024, AMA_FPL_2024, ARIES_FPGA_2025, RSN_ISCA_2025, GAMA_FPL_2025}.
The Versal SoC devices integrate an FPGA fabric, which acts as another level of memory hierarchy and is utilized for efficient exploitation of data reuse 
in GEMM.
Furthermore, the flexibility of the FPGA allows the design of customized tiling schemes \cite{Versal_vs_Stratix_FCCM_2024, CHARM_FPGA_2023} and tailored data layout transformations \cite{RSN_ISCA_2025}.

In this work, we conduct a comprehensive study to optimize GEMM workloads on Ryzen AI NPUs.
Due to the absence of FPGA fabric, GEMM mapping and optimization on the AMD NPUs is a fundamentally different problem compared to Versal devices.
This highlights the necessity for a separate methodology to address several new design challenges introduced by the NPUs.
Having identified this important research gap, we propose a \textit{unified} systematic framework to maximize GEMM performance across the two current generations, namely XDNA and XDNA2.
We extensively focus on the end-to-end GEMM performance  
and introduce a novel procedure to enable system-level optimization.
Within the framework, matrices are retained in regular order (\textit{row-} and \textit{column-major}) in main memory (DRAM).
This facilitates seamless integration with tensor libraries for DL, such as GGML \cite{GGML_github}, while also enabling the implementation of high-performance GEMM libraries, similar to GPUs \cite{cublas_doc, cutlass_doc, hipBLAS_doc, rocBLAS_doc}.
Moreover, our GEMM designs efficiently leverage several architectural features of the NPUs (\emph{e.g.,} on-the-fly tensor transformations), while also addressing critical performance bottlenecks (\emph{e.g.,} via sophisticated data movement design between the NPU and DRAM). The key contributions of this paper are:

\begin{itemize}
  \item  A systematic methodology to optimize GEMM workloads on Ryzen AI NPUs through analytical modeling along with hardware profiling. 
  Our methodology is general in scope; in this paper we apply it across the two current NPU generations. Observing the \textit{inverse} relationship between compute and off-chip memory, our approach is based on identifying the \textit{balanced} point that maximizes performance.
  
  \item A sophisticated GEMM implementation that enables sufficient contiguous DRAM accesses to maximize performance. 
  Our design leverages the multi-dimensional tensor addressing feature across the \textit{entire} NPU hierarchy to transform matrices into the tiled layout expected by NPU cores, enabling matrices to remain in standard layout in DRAM (\emph{i.e.,} \textit{row-} and \textit{column-} major order) without explicit pre-tiling.
  
  \item A thorough experimental evaluation on two mini PCs, each incorporating an NPU from a different generation, which exhibits GEMM performance up to 6.76 TOPS (XDNA) and 38.05 TOPS (XDNA2) for 8-bit integer (int8), as well as 3.14 TOPS (XDNA) and 14.71 TOPS (XDNA2) for brain floating-point (bf16).
  Furthermore, we present roofline sweeps to provide a holistic view of performance across hundreds of GEMM sizes for each data type.

    \item We provide essential insights regarding key performance considerations for GEMM optimization on AMD's NPU architecture.

  
\end{itemize}







\section{Related Work}
\label{sec:Related_work}

Having identified the importance of GEMM in DL, several prior works have proposed frameworks that aim to optimize GEMM workloads on Versal devices.
Some prior works \cite{CHARM_FPGA_2023, AutoMM_DAC_2023, Charm_v2_2024, ARIES_FPGA_2025, RSN_ISCA_2025} explore end-to-end GEMM performance including DRAM, while others \cite{MaxEVA_2023,  Versal_vs_Stratix_FCCM_2024, AMA_FPL_2024, GAMA_FPL_2025} focus exclusively on the AIE computation part (\emph{i.e.,} excluding off-chip considerations).
One common design choice across all of these works is to partition GEMM across the reduction dimension, driven by the limited number of ports in the AIE-FPGA interface.
For instance, MaxEVA \cite{MaxEVA_2023, Versal_vs_Stratix_FCCM_2024} utilizes 20\% of the AIE cores to perform adder-tree reduction, limiting the GEMM compute  efficiency to 80\%. 
In a similar manner, GAMA \cite{GAMA_FPL_2025} exploits the cascade interface to transfer partial accumulations across AIE cores, observing an average performance degradation of 7\% due to cascade stalls.
In contrast, as we demonstrate in this work, all cores on AMD NPUs can perform GEMM computation in an \textit{independent} fashion, thereby maximizing the compute efficiency.


Some prior works have also explored mapping GEMM on AMD NPUs. In \cite{roesti2025unlocking}, the authors explore fine-tuning GPT-2 on the Ryzen AI processors.
Specifically, they offload the time-intensive GEMM computation on XDNA and attain 2.8$\times$ speedup compared to a CPU-only implementation.
Furthermore, they utilize a GEMM implementation similar to the non-optimized programming example in \cite{GEMM_example_mlir_aie}.
In \cite{fang2025dato_paper}, a task-based programming model for dataflow accelerators is presented, demonstrating a GEMM implementation on XDNA. 
In particular, they attain up to 5.04 TOPS and 1.95 TOPS for int8 and bf16 data types, respectively, which closely matches the performance in \cite{GEMM_example_mlir_aie}.
In contrast, our optimized GEMM designs achieve superior performance of up to 6.76 TOPS (34\% higher) and 3.14 TOPS (62\% higher) on XDNA for int8 and bf16, respectively.

Related workloads were covered in prior publications.
The authors in \cite{ARIES_FPGA_2025} map a ResNet layer on XDNA, utilizing only 20\% of the XDNA cores. 
In \cite{Zen_attention_AMD_paper}, a compiler framework for dynamic attention folding on AMD NPUs is presented.
Moreover, in \cite{karami2025exploringdynamicschedulingspace} the authors perform characterization of generative AI workloads on Ryzen AI processors.
Finally, other works leverage the NPUs to perform stencil applications \cite{Stensil_NPU_FPGA2025}, accelerate Fortran intrinsics \cite{Fortran_instrinsics_FPGA2025}, and implement a variant of the fast Fourier transform (FFT) \cite{Variant_FFT_NPU_2024}.


\section{NPU Architecture, Features \& Programming}
\label{sec:NPU_architecture}

\begin{figure}[tbp]
\vspace{-0.30cm}
\centering
\includegraphics[width=0.74\linewidth]{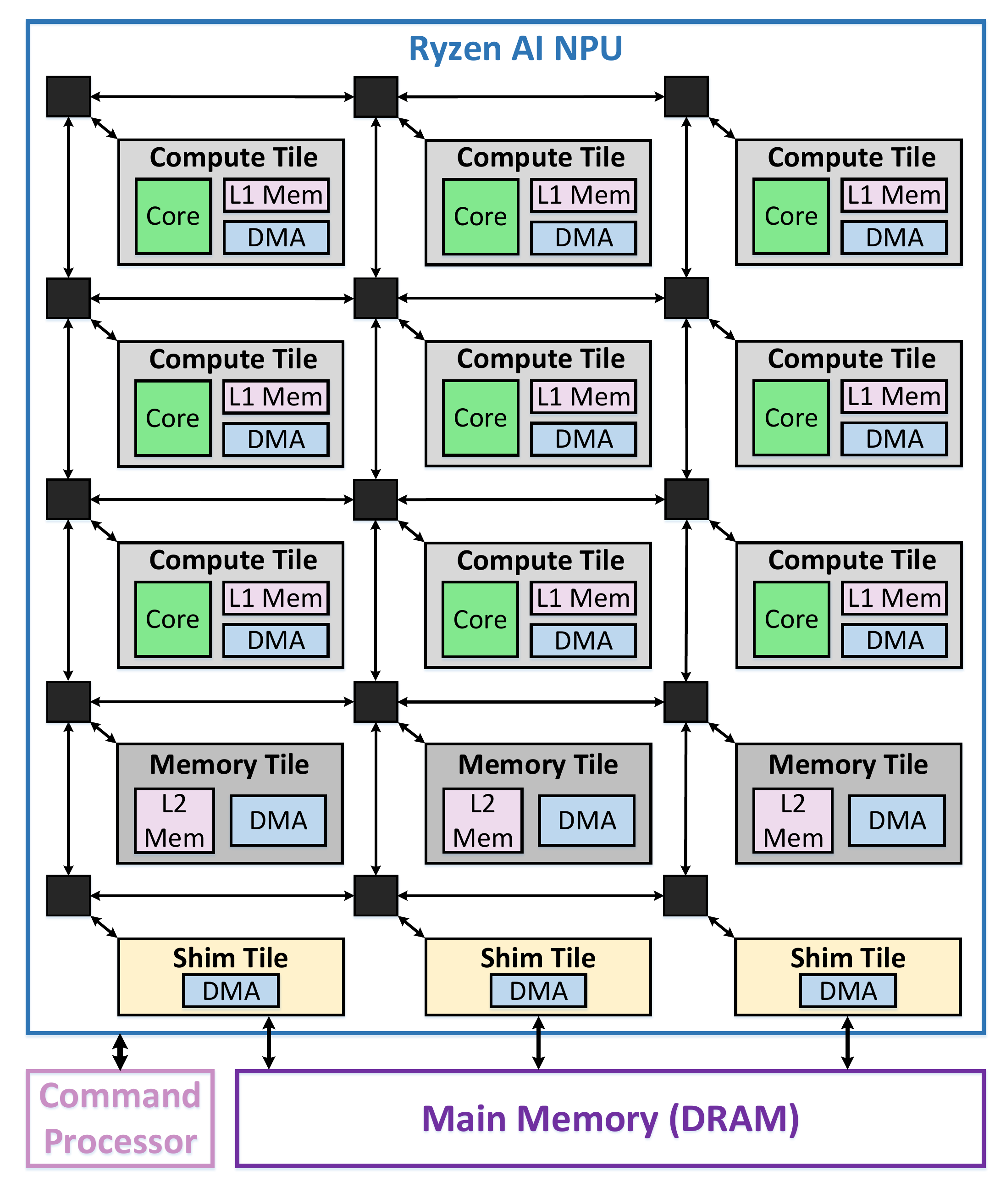}

\vspace{-0.40cm}

\caption{Architecture of Ryzen AI NPUs.}
\label{fig:NPU_architecture}

\vspace{-0.40cm}

\end{figure}

\subsection{NPU Architecture Overview}
\label{subsec:NPU_architecture}
The architecture of both XDNA and XDNA2 is illustrated in  Fig. \ref{fig:NPU_architecture}.
The NPU is a modular and scalable 
architecture, comprising a 2D array of identical compute tiles (CompTiles) \cite{NPU_arch_MICRO_2024, AMD_AIE_ML_architecture_manual}. 
CompTiles include the processing cores which operate out of a local memory (L1).
The NPU core is a very long instruction word (VLIW) processor with 
single instruction multiple data (SIMD) datapath, supporting both fixed-point and floating-point operations.
NPUs incorporate a second level of on-chip memory (L2) via the memory tiles (MemTiles).
These are arranged in a single row, located below the array of CompTiles (Fig. \ref{fig:NPU_architecture}).
Finally, NPUs include a last row of interface tiles (ShimTiles) to provide communication with DRAM.

Data movement between the levels of memory hierarchy is facilitated by direct memory access (DMA) engines, which are integrated across all NPU tiles.
DMAs move data between the NPU tiles by utilizing the configurable interconnects (switches), shown as black squares in Fig. \ref{fig:NPU_architecture}.
The DMA engines of ShimTiles read/write data to DRAM via the NPU network-on-chip (NoC) and SoC-level fabric of the Ryzen AI chips \cite{NPU_arch_MICRO_2024, NPU_HOT_CHIPS_2023}.
Task scheduling and data movement are orchestrated by an on-chip command processor.
This processor controls the data movement between ShimTiles and DRAM at runtime, allowing the NPU to focus exclusively on the main computation.
The command processor is also responsible for (re-)configuring the NPU compute kernels, switches, and DMA transfers.


The XDNA has 20 compute cores, which are organized as a 4$\times$5 array (rows $\times$ columns) of CompTiles \cite{NPU_arch_MICRO_2024}, while the XDNA2 contains 32 cores as a 4$\times$8 array \cite{IRON_MICRO_2024}.
Both NPU generations have 64 KB of memory per L1 tile and 512 KB of memory per L2 tile \cite{NPU_arch_MICRO_2024, AMD_AIE_ML_architecture_manual, Zen_attention_AMD_paper}.
XDNA and XDNA2 natively support int8, int16 and bf16 precisions.
Additionally, XDNA2 has hardware support for block floating-point (bfp16) datatype \cite{AMD_AIE_API_user_guide_2025.1}, where a block of eight numbers shares one common exponent \cite{AMD_quark_bfp16, Microsoft_block_floating_point_NEURIPS_2020}.
XDNA2 offers increased theoretical peak compute capabilities, delivering up to 50 TOPS \cite{AMD_CES, IRON_MICRO_2024}, compared to the 10 TOPS of XDNA \cite{NPU_arch_MICRO_2024}.




\subsection{Data Movement Architecture}
\label{subsec:data_movement_architecture}

\begin{figure}[t]
\vspace{-0.50cm}
\centering
\subfloat[]
{\includegraphics[width=0.35\linewidth]{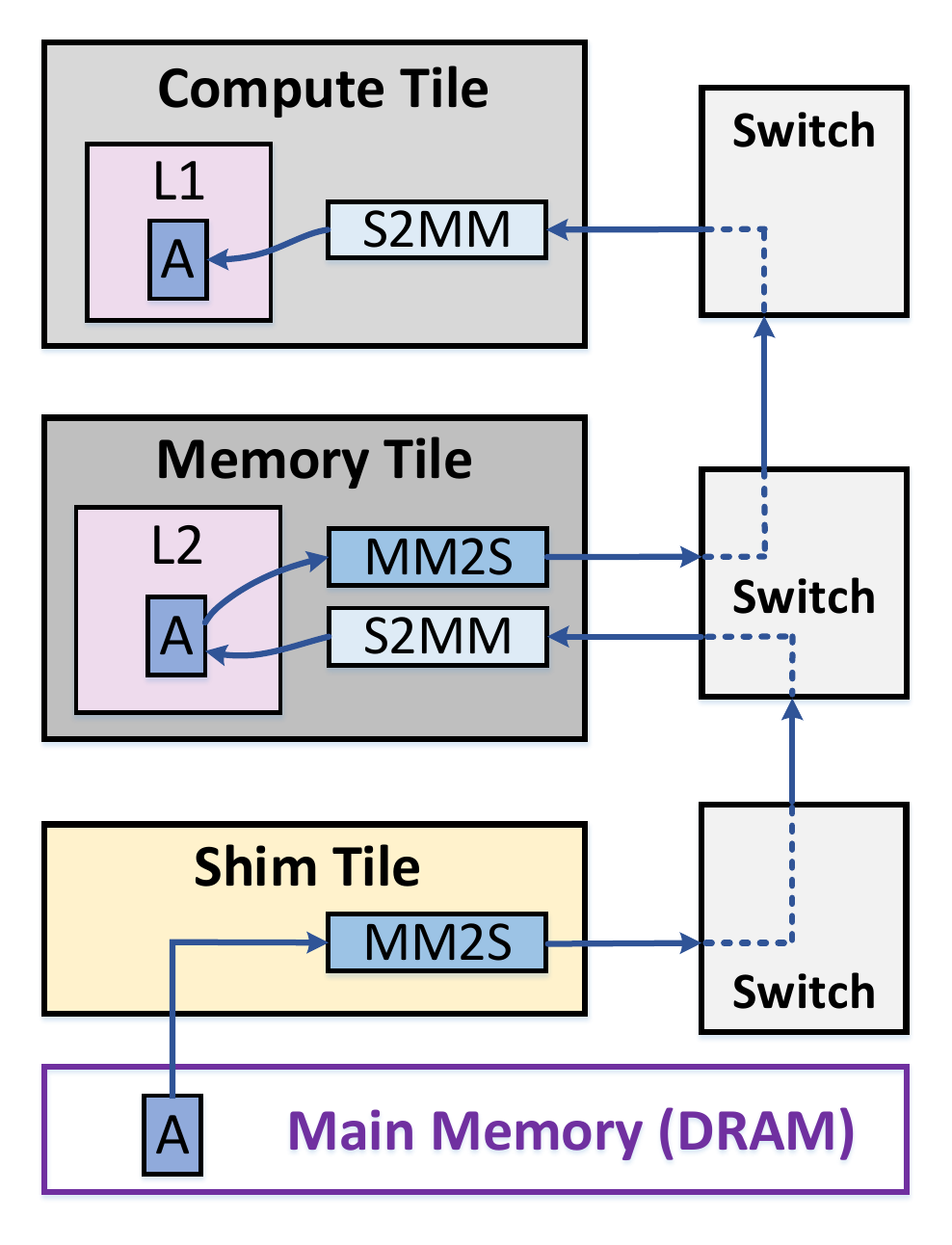}
\label{fig:Data_movement_matrix_a}}
\hspace{10mm}
\subfloat[]{\includegraphics[width=0.35\linewidth]{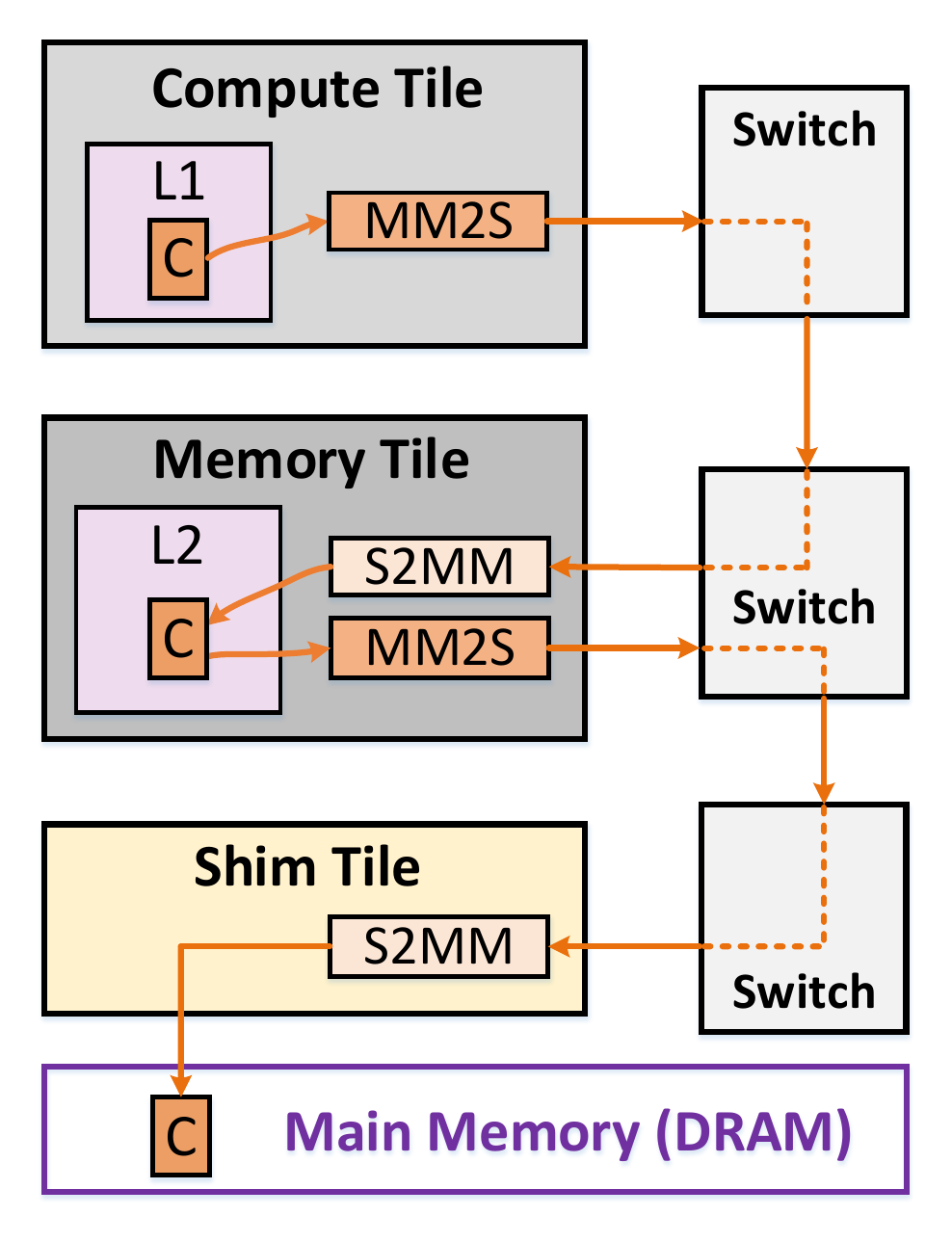}
\label{fig:Data_movement_matrix_c}} 

\vspace{-0.40cm}

\caption{Data movement across the memory hierarchy in Ryzen AI NPUs: input buffer A (a) and output buffer C (b).} 
\label{fig:data_movement_architecture}
\vspace{-0.40cm}
\end{figure}

Ryzen AI NPUs incorporate a dedicated data movement architecture that allows programmers to explicitly configure data transfers between all levels of the memory hierarchy. 
Data moves from a source DMA channel to a destination DMA channel through a circuit- or packet-switched stream, using one or more configurable switches between them.
The source channel is a memory-map to stream (MM2S) channel which reads data from memory and pushes it to the stream switches.
Conversely, the destination channel receives data from the stream switches and writes it to the memory-mapped space (S2MM).
Each CompTile and ShimTile includes two MM2S and two S2MM DMA channels, while MemTiles incorporate six MM2S and six S2MM channels \cite{AMD_AIE_ML_architecture_manual}.

Fig. \ref{fig:Data_movement_matrix_a} shows an example of moving an input buffer $A$ from DRAM to an L2 MemTile, and eventually to L1, where it can be utilized by the corresponding core.
In a similar fashion, Fig. \ref{fig:Data_movement_matrix_c} depicts the data movement of an output buffer from L1 to L2, and eventually to DRAM. 
The synchronization of data buffers between the DMAs and the corresponding module (\emph{e.g.,} the NPU core or DRAM) is managed by hardware lock units \cite{AMD_AIE_ML_architecture_manual, NPU_arch_MICRO_2024}.

DMAs run independently and in parallel with computation on cores.
They are programmed by configuring a sequence of  buffer descriptors (BDs).
A BD contains all the required information associated with a specific DMA transfer, such as the amount of data to read/write, the memory addresses involved, and the locks to acquire/release before and after each transfer \cite{AMD_AIE_ML_architecture_manual, AMD_AIE_ML_kernel_guide}.
BDs support both linear memory addressing and multi-dimensional address generation, enabling on-the-fly data layout transformations required for DL tensors.
CompTiles and ShimTiles support each 3D tensor addressing, while MemTiles incorporate 4D addressing.
This important DMA addressing feature is extensively exploited by our GEMM implementation across all tiles in the NPU architecture (refer to Sec. \ref{subsec:DMA_transformations}), allowing tensors to be stored in regular order in DRAM.

\subsection{Programming Tools}
\label{subsec:Programming_tools}

In this work, we use IRON, an open-source close-to-metal toolchain for developing programs on Ryzen AI NPUs \cite{mlir_aie_github_repo}.
As a low-level toolkit, IRON enables fine-grained control of the NPU architectural attributes, such as explicit data movement and complex access patterns supported in DMAs, while also providing convenient programming abstractions \cite{Erika_IRON_FCCM_2025}.
These characteristics render IRON a compelling choice for GEMM implementation, where explicit control of the  NPU architectural features is critical to performance.

\begin{figure*}[tbp]
\vspace{-0.50cm}
\centering
\includegraphics[width=1.00\textwidth]{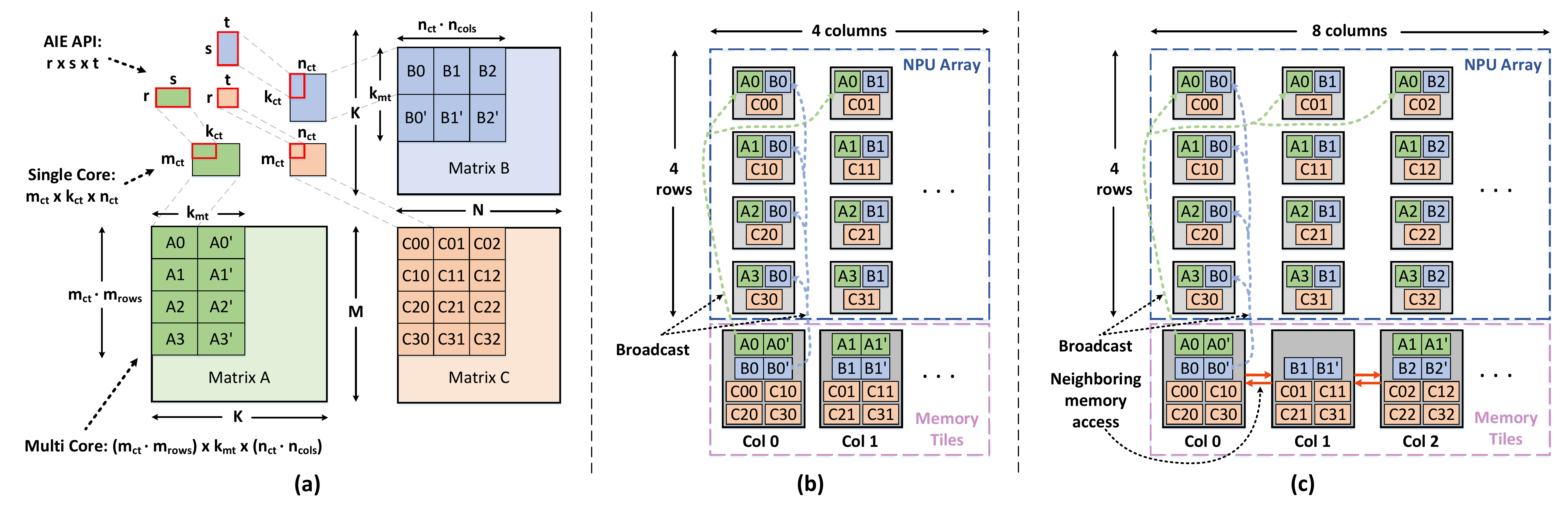}

\vspace{-0.40cm}

\caption{Proposed GEMM multi-level tiling scheme (a), and GEMM mapping strategy on XDNA (b) and XDNA2 (c).}
\label{fig:GEMM_tiling_mapping}
\vspace{-0.30cm}
\end{figure*}

IRON is based on a multi-level intermediate representation (MLIR) \cite{mlir_paper} dialect, named ``\textit{AIE}'', and offers a user-friendly interface to this dialect through Python bindings \cite{mlir_python_bindings}.
Hence, Python scripts can be used to generate code that describes both the data movement across the NPU hierarchy and the compute kernels that run on the cores.
The single-core compute kernels can be written in high-level C++ \cite{AMD_AIE_API_user_guide_2025.1}, or low-level SIMD intrinsics \cite{AIE_intrinsics}.
Finally, two options are available for single-core kernel compilation: the proprietary \emph{xchesscc} compiler \cite{AMD_AIE_ML_kernel_guide, mlir_aie_github_repo} and the open-source \emph{Peano} tool \cite{peano_github_repo}.



\section{GEMM Design \& Optimization}
\label{sec:GEMM_Design}

\subsection{Multi-Level Tiling Method}
\label{subsec:GEMM_tiling}

Fig. \ref{fig:GEMM_tiling_mapping}a illustrates the multi-level tiling scheme we use to partition the input and output matrices.
The inner-most (first) tiling level is defined by the supported shapes of the AIE API library for the single-core GEMM kernel \cite{AMD_AIE_API_user_guide_2025.1}, and is expressed via the parameters $r$ $\times$ $s$ $\times$ $t$.
The AIE API provides multiple optimized modes for each supported precision and enables portability across NPU generations.
The second tiling level is determined by the GEMM kernel that is supported out of local L1 memory, denoted as $m_\text{ct}$ $\times$ $k_\text{ct}$ $\times$ $n_\text{ct}$.
Since the NPU is a multi-core architecture, we partition GEMM across multiple cores to exploit spatial-level parallelism and maximize performance.
The third tiling level achieves this partitioning and corresponds to the GEMM size operating on the entire NPU array (parameters and mapping delineated in the next section).
Finally, the outer-most (fourth) tiling level is dictated by the final GEMM sizes of input matrices $A$ (typically \textit{activations}) and $B$ (typically \textit{weights}), and output matrix $C$, expressed as $M$ $\times$ $K$ $\times$ $N$.

\subsection{GEMM Mapping Strategy on NPUs}
\label{subsec:GEMM_mapping_strategy}


\subsubsection{\textbf{NPU Array Mapping \& Core Design}} 
\label{subsubsec:NPU_array_mapping}

Our mapping strategy is to parallelize GEMM in \textit{space} across the $M$ and $N$ dimensions, while reduction across the $K$ dimension is performed in \textit{time}.
In this manner, all NPU cores perform the \textit{same} GEMM computation on different data and operate \textit{independently}.
This leads to maximized performance, since no data communication occurs between the cores, as opposed to Versal devices, where the reduction dimension is partitioned across multiple cores (\emph{e.g.,} \cite{MaxEVA_2023, GAMA_FPL_2025}).
By using the broadcast feature of the NPU architecture, we can attain spatial parallelization and exploit the inherent data reuse of the GEMM algorithm.
This parallelization is expressed via the number of single-core tiles in the $M$ and $N$ dimensions, and is defined by the parameters $m_\text{rows}$ and $n_\text{cols}$, respectively (Fig \ref{fig:GEMM_tiling_mapping}a). 
In this work, we map the parameters $m_\text{rows}$ and $n_\text{cols}$ in a straightforward fashion to the number of rows and columns of the two NPU architectures, respectively.
In particular, we broadcast each input $A$ tile across one row in the NPU array (\emph{e.g.,} $A0$ in Fig. \ref{fig:GEMM_tiling_mapping}b and Fig. \ref{fig:GEMM_tiling_mapping}c for XDNA and XDNA2, respectively).
Similarly, each $B$ tile is broadcast across one column of NPU cores. 
Note that due to the absence of a ShimTile in the last column of XDNA, we choose to map GEMM across 4 rows and 4 columns for this architecture, similar to \cite{GEMM_example_mlir_aie, roesti2025unlocking, fang2025dato_paper}.
This enables a \textit{symmetric} 4 $\times$ 4 ($m_\text{rows}$ $\times$ $n_\text{cols}$) GEMM implementation, as depicted in Fig. \ref{fig:GEMM_tiling_mapping}b.
For XDNA2, we utilize the entire 4 $\times$ 8 array, which results in an \textit{asymmetric} mapping (Fig. \ref{fig:GEMM_tiling_mapping}c).





Each core performs a GEMM of $m_\text{ct}$ $\times$ $k_\text{ct}$ $\times$ $n_\text{ct}$ size.
To handle the reduction across $K$, the single-core kernel also loads previous partial results of $m_\text{ct}$ $\times$ $n_\text{ct}$ size, performs accumulation and stores the updated results back to the output tiles. 
For example, the upper-left core in Fig. \ref{fig:GEMM_tiling_mapping}b, sequentially loads tiles $A0$ and $B0$, $A0'$ and $B0'$, etc., in order to perform reduction.
In this fashion, the output $C$ tiles remain stationary in the L1 memory of each core (output stationary mapping).
When all tiles complete their accumulation ($K$/$k_\text{ct}$ tiles in total), the output tile is transferred to an L2 MemTile, and eventually to DRAM. The MemTile design is detailed in Sec. \ref{subsubsec:MemTiles_design}.
Subsequently, a fast vectorized kernel initializes the output $C$ tile to zero, preparing it for the next accumulation in GEMM tiling.

To overlap GEMM computation with DMA transfers of matrix tiles, we employ double-buffering on both L1 and L2 for the input tiles $A$ and $B$.
However, we retain the output $C$ tiles as a single buffer on each core.
Since the output tiles are transferred only once for each complete reduction, when $K$/$k_\text{ct}$ is sufficiently large, the infrequent added latency of a single-buffered output transfer becomes negligible.
At the same time, this design choice frees up valuable L1 memory, thereby enabling higher flexibility in tiling parameter optimization.
The larger tile sizes enabled by this additional memory ultimately lead to higher end-to-end GEMM performance in the general case (refer to Sec. \ref{subsubsec:single_output_buffer}).



\subsubsection{\textbf{Memory Tile Design}}
\label{subsubsec:MemTiles_design}

The input tiles are temporarily stored in L2 MemTiles before being broadcast to each NPU core.
Note that we load multiple input tiles across the reduction dimension $K$ into MemTiles (\emph{e.g.,} $A0$ and $A0'$ in Fig. \ref{fig:GEMM_tiling_mapping}b and \ref{fig:GEMM_tiling_mapping}c).
This is an important aspect of the mapping; 
loading multiple tiles allows us to access a larger amount of \textit{contiguous} data from DRAM, where matrices are stored in regular order (\textit{row-} and \textit{column-major}).
Such long contiguous reads result in increased DRAM bandwidth (BW) utilization, which is essential for maximizing 
GEMM performance at the system level.
In this work, we retain matrices $A$ and $C$ in \textit{row-major} order, while matrix $B$ is either in \textit{row-} or \textit{column-major} order.
To this end, we introduce another parameter, $k_\text{mt}$, which specifies the size of the tiles loaded into L2.
In particular, for matrix $A$, each L2 MemTile loads a tile of $m_\text{ct}$ $\times$ $k_\text{mt}$ size.
Similarly, when $B$ is in \textit{column-major} order, each L2 MemTile loads a tile of $k_\text{mt}$ $\times$ $n_\text{ct}$ size.
Storing matrix $A$ in \textit{row-major} and $B$ in \textit{column-major} provides sufficient contiguous DRAM access for both matrices, which leads to higher GEMM performance (Sec. \ref{subsubsec:GEMM_sweeps}).
However, when $B$ is in \textit{row-major}, MemTiles load the same tile as CompTiles (\emph{i.e.,} $k_\text{ct}$ $\times$ $n_\text{ct}$), since contiguous 
data are accessible across the $n_\text{ct}$ dimension. 


Due to the 4 $\times$ 4 symmetric design of XDNA, each MemTile holds the same amount of input tiles, as illustrated in Fig. \ref{fig:GEMM_tiling_mapping}b.
Specifically, each MemTile broadcasts $B$ tiles within its own column.
The $A$ tiles are broadcast across the four rows; we map them in a regular fashion to the four MemTiles: the MemTile in column 0 holds $A0$ (which will be broadcast across row 0), the MemTile in column 1 holds $A1$ (which will be broadcast across row 1), etc.
In contrast, for XDNA2, we map the four $A$ tiles to eight MemTiles in an alternating pattern across even columns due to its 4 $\times$ 8 asymmetric design.
The mapping for $B$ remains the same.
In this design, even MemTiles hold more tiles than their odd counterparts, as depicted in the \textit{logical} view of Fig. \ref{fig:GEMM_tiling_mapping}c.
This particular mapping aids the IRON tool to leverage the NPU architectural feature of directly accessing the memory of the neighboring MemTile \cite{AMD_AIE_ML_architecture_manual}.
Therefore, when buffer sizes exceed the capacity of a specific MemTile, IRON \textit{physically} allocates buffers to a neighboring MemTile.

Observe that four output $C$ tiles are aggregated across each column into a MemTile (Fig. \ref{fig:GEMM_tiling_mapping}b and \ref{fig:GEMM_tiling_mapping}c).
This is because four $C$ tiles need to be transferred concurrently, while ShimTiles provide only two S2MM channels.
Hence, we exploit the six S2MM channels available in each MemTile \cite{AMD_AIE_ML_architecture_manual} to temporarily store the four output tiles, before they are transferred to DRAM via ShimTiles.

We define the \textit{native} GEMM size, as ($m_\text{ct}$ $\cdot$ $m_\text{rows}$) $\times$ $k_\text{mt}$ $\times$ ($n_\text{ct}$ $\cdot$ $n_\text{cols}$).
This corresponds to the GEMM size that operates natively on the entire NPU array, while also ensuring high performance.
Moreover, note that although we arbitrarily map matrix $A$ across rows and matrix $B$ across columns, the reverse mapping is equally feasible, yielding symmetrical solutions across the $M$ and $N$ dimensions.
Finally, although an input stationary mapping can also be employed, it would not be adequate to efficiently support arbitrary GEMM dimensions. 
Specifically, partial results would need to be temporarily stored in MemTiles, and subsequently reloaded to CompTiles for exploitation of data reuse in GEMM.
This would require three input channels for efficient GEMM computation, while CompTiles provide two inputs channels.




\subsection{On-The-Fly Tensor Transformations}
\label{subsec:DMA_transformations}

\begin{figure}[tbp]
\vspace{-0.40cm}
\centering
\includegraphics[width=0.70\linewidth]{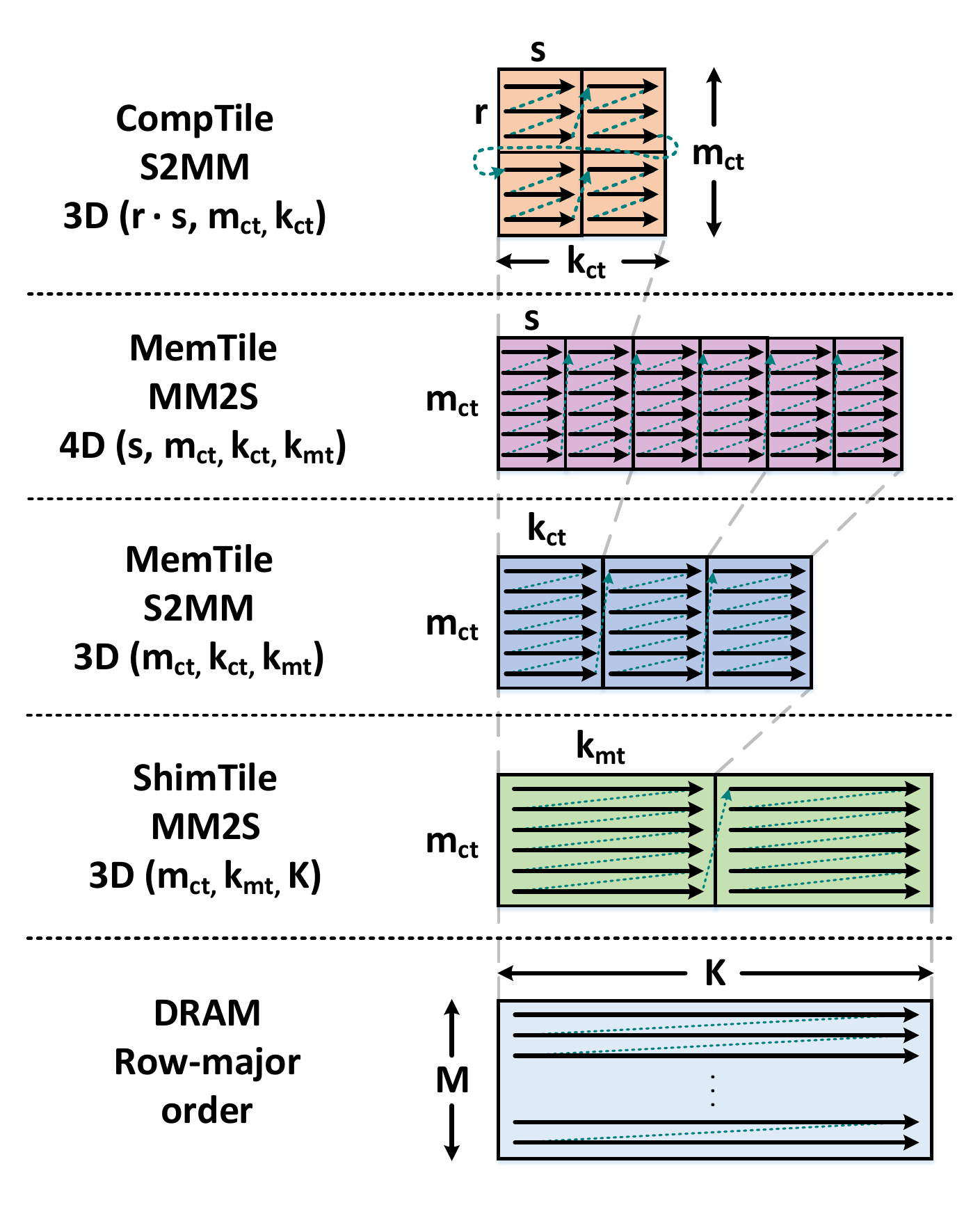}

\vspace{-0.40cm}

\caption{On-the-fly DMA transformations for matrix A.}
\label{fig:DMA_transformations_mat_A}

\vspace{-0.30cm}

\end{figure}

We extensively exploit the multi-dimensional addressing feature of DMAs 
to reorganize data into tiled layouts, as needed by the NPU cores.
This enables matrices to be stored in standard order in DRAM (\textit{row-} and \textit{column-major}).
The single-core GEMM kernels assume that matrices are pre-tiled \cite{AMD_AIE_API_user_guide_2025.1}.
In particular, the kernels running on each core expect $r\times s \times t$-sized tiles and both data \textit{within} tiles as well as the tiles \textit{themselves} to be in \textit{row-major} order, as illustrated in the upper part of Fig. \ref{fig:DMA_transformations_mat_A}, for the case of matrix $A$.
The aforementioned distribution of tiles across multiple cores, along with the requirement for contiguous DRAM accesses necessitates multiple data layout transformations, as explained below.

As matrices are transferred from DRAM to NPU, DMAs apply a series of transformations depending on the addressing capabilities of each NPU tile. 
Fig. \ref{fig:DMA_transformations_mat_A} shows the transformations of each DMA channel associated with the transfer of matrix $A$ tiles (see Fig. \ref{fig:Data_movement_matrix_a} for DMA channels).
Initially, a ShimTile MM2S channel reads a tile of $m_\text{ct}$ $\times$ $K$ size from DRAM. 
Since ShimTiles support 3D addressing, the \textit{row-major} $m_\text{ct}$ $\times$ $K$ tile is transformed into multiple smaller $m_\text{ct}$ $\times$ $k_\text{mt}$  tiles via the MM2S channel (parameters: $m_\text{ct}$, $k_\text{mt}$, $K$).
Before each $m_\text{ct}$ $\times$ $k_\text{mt}$ tile is stored in the MemTile, another 3D transformation occurs at the S2MM MemTile channel, partitioning it into several $m_\text{ct}$ $\times$ $k_\text{ct}$ tiles (parameters: $m_\text{ct}$, $k_\text{ct}$, $k_\text{mt}$).

Each MemTile holds a $m_\text{ct}$ $\times$ $k_\text{mt}$ tile, which it sequentially transmits to the corresponding CompTiles as a series of smaller $m_\text{ct}$ $\times$ $k_\text{ct}$ tiles.
Since the CompTiles expect pre-tiled data, the data layout of each of the smaller tiles requires transformation.
Describing this transfer requires five parameters, namely $r$, $s$, $m_\text{ct}$, $k_\text{ct}$, $k_\text{mt}$.
However, MemTiles only support 4D addressing. 
To circumvent this, we decompose the transformation into two separate transformations, by utilizing the hardware's data layout transformation features in both the MM2S MemTile (output) channel and the S2MM CompTile (input) channel.
First, the MM2S MemTile channel partitions data into several $m_\text{ct}$ $\times$ $s$ tiles, as depicted in Fig. \ref{fig:DMA_transformations_mat_A} (parameters: $s$, $m_\text{ct}$, $k_\text{ct}$, $k_\text{mt}$).
This enables address \textit{linearization} within the $r$ $\times$ $s$ tile, thereby allowing a subsequent 3D transformation in the S2MM CompTile channel to reorganize data into the required layout
(\textit{effective} parameters: $r \cdot s$, $m_\text{ct}$, $k_\text{ct}$).

Address generation in DMAs occurs at 32-bit granularity \cite{AMD_AIE_ML_architecture_manual, AMD_AIE_ML_kernel_guide}.
DMAs alone cannot perform layout transformations at smaller-precision data types (\textit{e.g.,} individual elements of int8 or bf16 data types).
However, shuffle instructions running on the AIE cores can be utilized to support this operation and enable the fine-grained data swizzling.
In our application, this becomes relevant in use cases where an int8 or bf16 matrix $B$ is stored in \textit{column-major} order in DRAM.
To this end, we modify the GEMM kernel to utilize shuffling instructions, by using the AIE API transpose function \cite{transpose_AIE_API}, such that both data \textit{within} tiles and the tiles themselves are in \textit{column-major} order. 
Subsequently, we apply similar transformations to matrix $B$ as demonstrated for matrix $A$, across the NPU hierarchy (Fig. \ref{fig:DMA_transformations_mat_A}).

Furthermore, when matrix $B$ is in \textit{row-major}, only one 4D transformation is required in the MemTiles (parameters: $s$, $t$, $k_\text{ct}$, $n_\text{ct}$), since each MemTile holds a $k_\text{ct}$ $\times$ $n_\text{ct}$ tile (Sec. \ref{subsec:GEMM_mapping_strategy}).
In a similar fashion, matrix $C$ tiles (\textit{row-major}) entail a single 4D transformation in the MemTiles (parameters: $r$, $t$, $m_\text{ct}$, $n_\text{ct}$).

\begin{figure}[tbp]
\vspace{-0.20cm}
\centering
\includegraphics[width=0.78\linewidth]{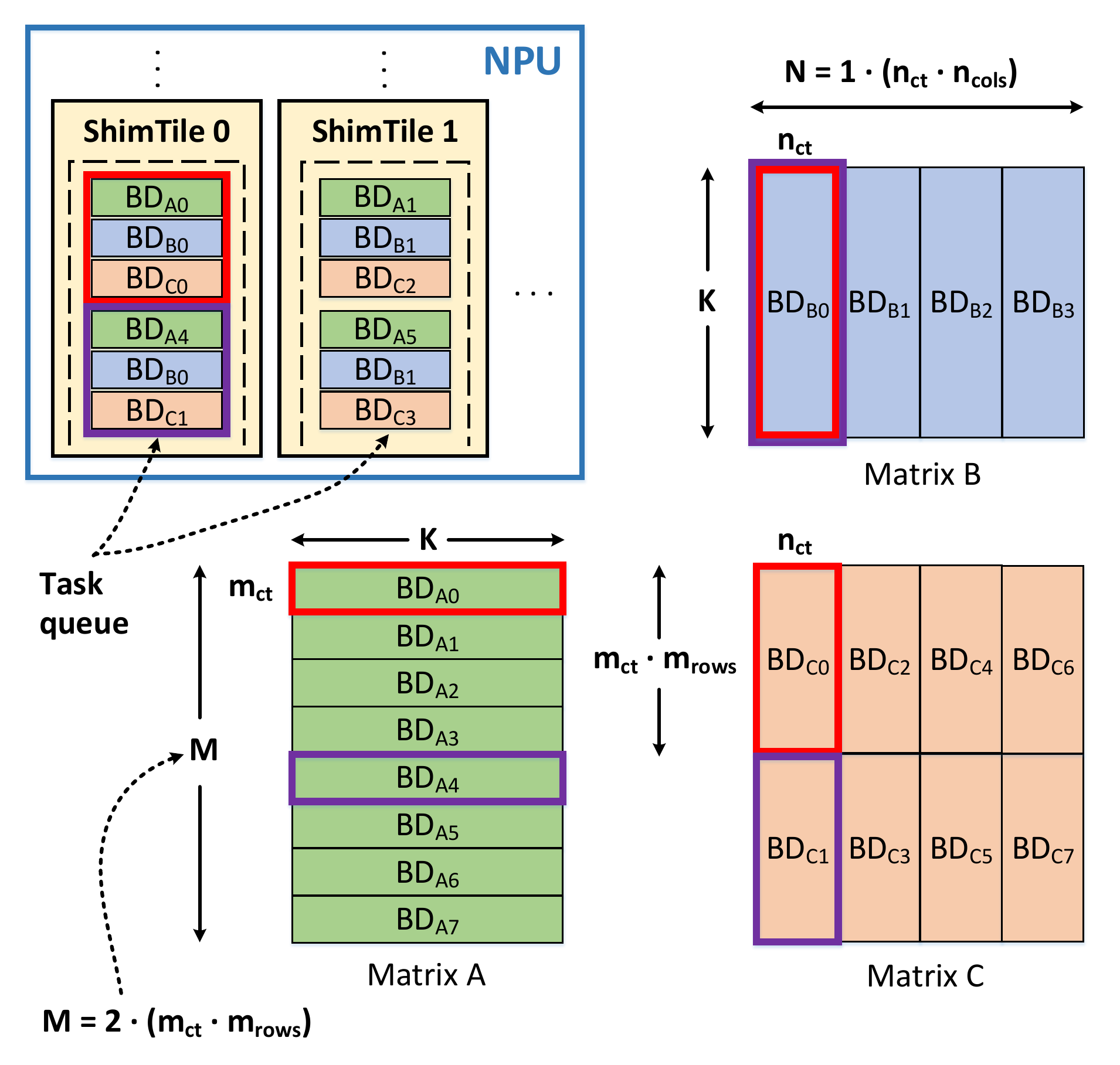}

\vspace{-0.40cm}

\caption{Simplified view of outer-most (fourth) GEMM tiling level, determined by NPU--DRAM transfers ($m_\text{rows}$, $n_\text{cols}=4$).}
\label{fig:DRAM_NPU_data_movement}

\vspace{-0.40cm}

\end{figure}

\subsection{Data Movement Between NPU \& DRAM}
\label{subsec:npu_main_memory_data_movement}

The outer-most (fourth) level of GEMM tiling encompasses data movement between the NPU and DRAM. 
The NPU interfaces with DRAM via ShimTiles and the control of data movement is orchestrated by the on-chip command processor.
ShimTiles are equipped with an input task queue to facilitate DMA transfers, where tasks are submitted sequentially.
When a task terminates, it issues a task-completion token, which the command processor uses to synchronize between multiple DRAM transfers.
Each ShimTile has access to 16 BDs \cite{AMD_AIE_ML_architecture_manual}, which are used to specify DMA transfers.
When a complex data movement pattern requires more than 16 BDs on a ShimTile, we can reuse (reconfigure) BDs. 
Before reconfiguring a BD, it is necessary to properly synchronize and ensure that the previous transfer associated with the BD has completed \cite{runtime_data_movement_mlir_aie}.

Fig. \ref{fig:DRAM_NPU_data_movement} illustrates a simplified view of the data movement between the NPU and DRAM.
Each BD is utilized to describe data movements in a fine-grained manner.
In particular, for matrix $A$, each BD defines one ShimTile DMA transfer of $m_\text{ct}$ $\times$ $K$ size.
Similarly, for matrix $B$, each BD defines a transfer of $K$ $\times$ $n_\text{ct}$. 
For matrix $C$, each BD describes a transfer of ($m_\text{ct}$ $\cdot$ $m_\text{rows}$) $\times$ $n_\text{ct}$, since $m_\text{rows}$ output tiles are aggregated per column.
The command processor program in our implementation maps and inserts BDs into ShimTile task queues as determined by the GEMM mapping strategy (\emph{e.g.,} $BD_{A0}$,  $BD_{B0}$, and $BD_{C0}$ to ShimTile 0, $BD_{A1}$, $BD_{B1}$, and $BD_{C2}$ to ShimTile 1, etc).
Moreover, $BD_{A4}$, $BD_{B0}$, and $BD_{C1}$ are also pushed into the queue of ShimTile 0, as dictated by GEMM tiling.
Similarly, our program enqueues BDs describing the GEMM transfers into the input task queues of each ShimTile, in a sequential fashion.
We note that the simplified example in Fig. \ref{fig:DRAM_NPU_data_movement} is directly applicable to XDNA ($n_\text{cols}=4$, see Sec. \ref{subsec:GEMM_mapping_strategy}), while the BD mapping for XDNA2 can be determined in straightforward fashion.

Depending on the target GEMM dimensions, (\emph{i.e.,} $M$, $K$, $N$), more than the maximum of 16 BDs in each ShimTile might be required.
BD reconfiguration is needed in this situation, which might result in performance degradation (Sec. \ref{subsubsec:BD_reconfiguration}).
To address this challenge, we propose the following procedure, which enables DMA data transfers to overlap with BD reconfiguration.
Initially, we submit to the task queue five BDs for each of the three $A$, $B$, and $C$ DMA transfers.
This efficiently utilizes 15 out of the 16 BDs available in each ShimTile.
Each DMA transfer begins immediately once its associated BD reaches the front of the queue.
For instance, $BD_{A0}$ and $BD_{B0}$ transfers start first in Fig \ref{fig:DRAM_NPU_data_movement}.
Subsequently, the command processor waits for a task-completion token for each output transfer in a sequential manner (\emph{e.g.,} first for $BD_{C0}$, then for $BD_{C1}$, etc).
Notice that the command processor only needs to wait for completion of each BD associated with the output matrix (\emph{e.g.,} $BD_{C0}$), since once it completes, the corresponding input BDs (\emph{e.g.,} $BD_{A0}$ and $BD_{B0}$) have also finished.
Therefore, once each output BD completes, the three retired BDs can be safely reconfigured, and the next three BDs are inserted into the queue (if available).
This ensures that 15 BDs are in the queue in the steady-state operation, allowing DMA data movement to overlap efficiently with BD reconfiguration.
This process is repeated iteratively until GEMM tiling is completed.

The fine-grained BD description of NPU--DRAM data movement discussed above allows supporting very large GEMM dimensions.
GEMM dimensionality is limited by the NPU registers bitwidth utilized in multi-dimensional tensor addressing \cite{AMD_AIE_ML_kernel_guide}.
For instance, when storing $B$ in \textit{column-major}, the GEMM programming example in \cite{GEMM_example_mlir_aie} allows only a reduction $K$ dimension of up to $\sim$4K for bf16 on XDNA2, while our approach allows sizes $>$ 64K in all dimensions.



\subsection{Analytical Modeling Optimization}
\label{subsec:analytical_modeling_optimization}

To maximize GEMM performance, we propose an optimization methodology based on analytical modeling.
First, we focus on single-core performance to gain insights for the on-chip compute part, 
and then extend our methodology to optimize the system-level performance by incorporating off-chip DRAM BW constraints.

\subsubsection{\textbf{Single-Core GEMM Optimization}} 
\label{subsubsec:singe_core_optimization}

Our model utilizes architectural parameters, such as the peak compute throughput of the cores ($peak\_MACs$ in MACs/cycle) and the DMA BW ($DMA\_BW$ in Bytes/cycle), in order to identify the optimal $m_\text{ct}$, $k_\text{ct}$, $n_\text{ct}$ parameters that maximize performance.
We define the efficiency ($eff$) as the fraction of the attained compute throughput to the peak throughput of the core.
Eq. \ref{eq:compute_cycles} expresses the compute cycles of the GEMM kernel ($C_{comp}$), while Eq. \ref{eq:comm_A_cycles} and \ref{eq:comm_B_cycles} formulate the number of cycles needed for the DMA transfers of $A$ ($CA_{comm}$) and $B$ ($CB_{comm}$) tiles, respectively.
Here, $ty(\cdot)$ indicates the data type size (in Bytes) of the matrices.
\begin{gather}
C_{comp} = m_\text{ct} \cdot k_\text{ct} \cdot n_\text{ct}/ (eff \cdot peak\_MACs) \label{eq:compute_cycles}\\
CA_{comm} = m_\text{ct} \cdot k_\text{ct} \cdot ty(A)/ DMA\_BW \label{eq:comm_A_cycles}\\
CB_{comm} = k_\text{ct} \cdot n_\text{ct} \cdot ty(B)/ DMA\_BW
\label{eq:comm_B_cycles}
\end{gather}
Next, we define the constraint in Eq. \ref{eq:compute_bound_constr} to ensure that the single-core GEMM remains compute bound (not bounded by the DMA BW for $A$ and $B$ tiles).
Furthermore, Eq. \ref{eq:memory_constr} restricts the GEMM buffers to fit within the 64KB L1 memory capacity (with 1KB reserved for stack).
Finally, we impose the straightforward constraint that $m_\text{ct}$, $k_\text{ct}$, $n_\text{ct}$ need to be multiples of $r$, $s$, $t$, respectively (not shown).
\begin{gather}
C_{comp} \geq \{CA_{comm},\ CB_{comm}\} \label{eq:compute_bound_constr} \\
\begin{split}
\{2 \cdot m_\text{ct} \cdot k_\text{ct} \cdot ty(A) 
&+ 2 \cdot k_\text{ct} \cdot n_\text{ct} \cdot ty(B) \\
&+ m_\text{ct} \cdot n_\text{ct} \cdot ty(C)\} \leq 63\,\text{KB} \label{eq:memory_constr}
\end{split}
\end{gather}
The solution of $m_\text{ct}$, $k_\text{ct}$, $n_\text{ct}$ can be formulated as an integer programming (IP) optimization problem utilizing the aforementioned constraints.
The IP is solved exhaustively by setting the maximization of the number of MACs ($m_\text{ct}$ $\cdot$ $k_\text{ct}$ $\cdot$ $n_\text{ct}$) as the main objective.
This increases data reuse in GEMM, thereby maximizing the overall efficiency.
Furthermore, due to the output stationary GEMM mapping, we impose a second objective to minimize the output $C$ tile (\emph{i.e.,} the product $m_\text{ct}$ $\cdot$ $n_\text{ct}$).
This is essential in reducing the number of loads/stores for accumulations 
and decreasing memory stalls caused by bank conflicts.
Evidently, the two optimization objectives lead to increased $k_\text{ct}$ and reduced $m_\text{ct}$, $n_\text{ct}$ (sufficiently large to not become DMA BW bound).
Moreover, notice that we do not impose DMA constraints on the output $C$ buffer, as opposed to $A$ and $B$ (Eq. \ref{eq:comm_A_cycles}, \ref{eq:comm_B_cycles}), while also retaining $C$ as a single buffer (Eq. \ref{eq:memory_constr}). 
This significantly increases the search space, thus increasing performance in the \textit{general} GEMM case (Sec. \ref{subsubsec:single_output_buffer}), which is an essential aspect of the system-level optimization discussed below.


\subsubsection{\textbf{System-Level NPU Array Optimization}} 
\label{subsubsec:system_level_optimization}

First, we analytically express the DRAM accesses for each matrix in GEMM. 
Eq. \ref{eq:A_DRAM_footprint} captures the DRAM reads needed for matrix $A$ ($A_{mem}$).
The first term, $m_\text{ct} \cdot m_{\text{rows}} \cdot K \cdot ty(A)$, represents the DRAM reads (in Bytes) during GEMM tiling along the $K$ dimension.
This is due to the output stationary mapping and because  $A$ is broadcast across rows (Sec. \ref{subsec:GEMM_mapping_strategy}).
The second term, $N/(n_\text{ct} \cdot n_\text{cols})$, describes the repeat factor of the aforementioned read accesses due to tiling along the $N$ dimension. 
In a similar manner, the third term, $M/(m_\text{ct} \cdot m_\text{rows})$, captures the repeat across the $M$ dimension.
\begin{gather}
A_{mem} = 
\left(m_\text{ct} \cdot m_{\text{rows}} \cdot K \cdot ty(A)\right)
\left(\frac{N}{n_\text{ct} \cdot n_\text{cols}}\right)
\left(\frac{M}{m_\text{ct} \cdot m_\text{rows}}\right) \nonumber \\
\Rightarrow \
A_{mem} = M \cdot K \cdot N \cdot ty(A)/(n_\text{ct} \cdot n_\text{cols})
\label{eq:A_DRAM_footprint}
\end{gather}
Similarly, Eq. \ref{eq:B_DRAM_footprint} represents the DRAM reads for matrix $B$ ($B_{mem}$), while Eq. \ref{eq:C_DRAM_footprint} shows the DRAM writes for the output matrix $C$ ($C_{mem}$).
We note that these equations provide an elegant and compact representation of data reuse and tiling scheme in GEMM.
\begin{gather}
B_{mem} = M \cdot K \cdot N \cdot ty(B)/(m_\text{ct} \cdot m_\text{rows}) \label{eq:B_DRAM_footprint} \\
C_{mem} = M \cdot N \cdot ty(C)
\label{eq:C_DRAM_footprint}
\end{gather}
Furthermore, Eq. \ref{eq:NPU_time} models the GEMM compute time on the NPU ($T_{comp}$), while Eq. \ref{eq:DRAM_time} expresses the total DRAM access time ($T_{mem}$).
Here, $peak\_TOPS$ denotes the \textit{theoretical peak} throughput of the NPU array, calculated at the maximum operating frequency.
Furthermore, $DRAM\_BW$ is the \textit{effective} DRAM BW achieved during GEMM execution on the NPU.
Note that, since all NPU cores execute the same GEMM kernel \textit{independently}, the single-core efficiency $eff$, as defined in Sec. \ref{subsubsec:singe_core_optimization}, directly corresponds to the entire NPU array efficiency in Eq. \ref{eq:NPU_time}.
\begin{gather}
T_{comp} = 2 \cdot M \cdot K \cdot N /(eff \cdot peak\_TOPS) \label{eq:NPU_time} \\
T_{mem} = (A_{mem} + B_{mem} + C_{mem})/ DRAM\_BW
\label{eq:DRAM_time}
\end{gather}
As discussed previously (Sec. \ref{subsubsec:singe_core_optimization}), minimizing $m_\text{ct}$ and $n_\text{ct}$, as well as maximizing $k_\text{ct}$ is essential to attain high efficiency, and thus maximized GEMM performance (equivalent to minimized $T_{comp}$).
However, from Eq. \ref{eq:A_DRAM_footprint} and \ref{eq:B_DRAM_footprint}, we notice that $n_\text{ct}$ and $m_\text{ct}$ parameters appear on the denominator of DRAM accesses for $A$ and $B$, respectively.
Therefore, the DRAM access time, $T_{mem}$, increases as $m_\text{ct}$ and $n_\text{ct}$ decrease.
This highlights the \textit{inverse} relationship between compute and memory time: \textit{as one increases, the other decreases}.
Thus, the optimal GEMM performance is attained at the \textit{balanced} point where they intersect (\emph{i.e.,} $T_{comp}$ $\approx$ $T_{mem}$).



In order to find the aforementioned optimal \textit{balanced} point we empirically set starting values for $m_\text{ct}$, $k_\text{ct}$, $n_\text{ct}$, and $eff$ parameters, and perform the iterative procedure explained below (starting values reduce iterations to typically $<$5).
These starting values are set based on the results of the single-core kernel optimization (Sec. \ref{subsubsec:singe_core_optimization}), and the effective $DRAM\_BW$ during GEMM execution (measured via micro-benchmarking, refer to Sec. \ref{subsubsec:optimal_balanced_GEMM_kernel}).
First, we measure the actual GEMM performance on the NPU device as well as the single kernel efficiency for the starting values, verifying that GEMM is memory bound ($T_{comp}$ $<$ $T_{mem}$, due to low $m_\text{ct}$, $n_\text{ct}$, and high $k_\text{ct}$).
In each iteration, we decrease the parameter $k_\text{ct}$ (as a multiple of $s$), and solve exhaustively an IP similar to the previous Sec. (\ref{subsubsec:singe_core_optimization}).
However, for that specific iteration, we fix the $k_\text{ct}$ parameter and the objective is to maximize the product of $m_\text{ct}$ $\cdot$ $n_\text{ct}$.
This leads to maximized $m_\text{ct}$, $n_\text{ct}$ values, given the DMA BW and L1 memory constraints, ensuring a highly optimized GEMM kernel (maximized number of MACs).
This is essential in identifying the optimal \textit{balanced} point, because it results in the smallest possible increment of $T_{comp}$ in each iteration, while also ensuring maximized possible GEMM performance given the specific parameters of that iteration.
We note here that $eff$ is estimated based on each previous point measurements of each iteration.
We iteratively measure GEMM performance on the NPU device of the top-ranked solution for each IP, verifying that in each step performance is higher compared to the previous.
While GEMM performance is increasing in each step, at some point we observe lower performance, and the iteration stops.
Evidently, at this specific point GEMM has become compute bound ($T_{comp}$ $>$ $T_{mem}$), while also performance is lower.
Therefore, the \textit{balanced} point where GEMM performance is maximized has been identified ($m_\text{ct}$, $k_\text{ct}$, $n_\text{ct}$ parameters of the previous iteration).


\section{Evaluation}
\label{sec:Evaluation}

For the experimental evaluation, we use two representative mini PCs, corresponding to two NPU generations.
The Minisforum UM790 Pro \cite{Phoenix_mini_PC}, equipped with Ryzen 9 7940HS processor (\textit{Phoenix Point}) \cite{Phoenix_processor}, is used for XDNA.
For XDNA2, we use the ASRock 4$\times$4 Box \cite{Krackan_mini_PC}, featuring the AMD Ryzen AI 7 350 processor (\textit{Krackan Point}) \cite{Krackan_processor}.
Both mini PCs have dual-channel DDR5-5600 MT/s DRAM.
The CPUs run Ubuntu 24.04 LTS and serve as the host.
In particular, CPUs allocate 
buffers in DRAM using Xilinx Runtime (XRT) \cite{xrt_github_repo}, configure the NPU, invoke the NPU for GEMM execution, and process the completion notification.
Finally, throughout all experiments, NPUs are configured at their maximum performance level (\emph{i.e.,} \textit{turbo} mode \cite{xrt_smi_commands}, utilizing XDNA driver commands \cite{xdna_driver_github_repo}).

\begin{table}[t]
\vspace{-0.10cm}
\centering
\caption{Single-core GEMM results for XDNA \& XDNA2.}
\setlength\tabcolsep{3.8pt}
\renewcommand{\arraystretch}{0.98}
\resizebox{0.88\linewidth}{!}{
\begin{threeparttable}
\begin{tabular}{c|c|c|c|c}

\Xhline{2.5\arrayrulewidth}

\multirow{2}{*}{\rotatebox[]{90}{\textbf{Dev.}}} & \textbf{Precision}  &  \textbf{Kernel Size} & \textbf{Throughput}  & \textbf{L1 Core}\\

& \textbf{In-Out}  &  \textbf{$m_\text{ct}$ $\times$ $k_\text{ct}$ $\times$ $n_\text{ct}$} & \textbf{MACs/cycle} & \textbf{Mem. (KB)}\\

\hline
\hline

\multirow{4}{*}{\rotatebox[]{90}{\textbf{XDNA}}} & int8-int8 & 64 $\times$ 232 $\times$ 64 & 233.0 & 62.0 (97\%)\\ 

& int8-int16 & 64 $\times$ 216 $\times$ 64 & 217.6 & 62.0 (97\%)\\ 

& int8-int32 & 48 $\times$ 280 $\times$ 48 & 192.0 & 61.5 (96\%)\\ 

& bf16-bf16 & 64 $\times$ 104 $\times$ 64 & 112.6 & 60.0 (94\%)\\ 

\hline

\multirow{4}{*}{\rotatebox[]{90}{\textbf{XDNA2}}} & int8-int8 & 64 $\times$ 232 $\times$ 64 & 450.6 & 62.0 (97\%)\\ 

& int8-int16 & 64 $\times$ 216 $\times$ 64 & 419.8 & 62.0 (97\%)\\ 

& int8-int32 & 48 $\times$ 280 $\times$ 48 & 384.0 & 61.5 (96\%)\\ 

& bf16-bf16 & 48 $\times$ 152 $\times$ 48 & 158.1 &   61.5 (96\%)\\ 

\Xhline{2.5\arrayrulewidth}

\end{tabular}


\end{threeparttable}
}

\label{tb:single_core_GEMM_results}

\vspace{-0.40cm}

\end{table}

\subsection{Single-Core GEMM Performance}
\label{subsec:single_core_GEMM_performance}


We exploit the AIE API \cite{AMD_AIE_API_user_guide_2025.1} to design highly optimized single-core GEMM kernels.
The kernels are compiled using the \emph{xchesscc} tool, employing various compiler directives (\emph{e.g.,} software pipelining and loop unrolling/flattening) to attain high efficiency.
Performance is assessed via hardware profiling utilizing the NPU trace unit \cite{AMD_AIE_ML_architecture_manual, tracing_mlir_aie}, thereby enabling cycle accurate measurements.
For int8, besides full output precision (int32), we also perform precision reduction to 16- and 8-bits, which is a common technique to increase GEMM performance in AIE architectures \cite{Vitis_tutorials, Charm_v2_2024, AMA_FPL_2024, GAMA_FPL_2025}.
Moreover, for bf16, we retain the output precision to bf16.
Finally, for XDNA2 we emulate the bf16 precision utilizing the bfp16 hardware datapath, which leads to increased GEMM performance (emulation attained via a specific \emph{xchesscc} flag at compile time, as mentioned in \cite{AMD_AIE_API_user_guide_2025.1}).



In Table \ref{tb:single_core_GEMM_results}, we present the top-ranked solutions of the single-core optimization procedure described in Sec. \ref{subsubsec:singe_core_optimization}.
For int8 precision, we attain very high throughput, ranging from 192.0--233.0 MACs/cycle and from 384.0--450.6 MACs/cycle for XDNA and XDNA2, respectively. 
Similarly for bf16, we achieve 112.6 MACs/cycle for XDNA, and 158.1 MACs/cycle for XDNA2.
Observe that in both cases, XDNA2 attains higher throughput compared to XDNA, since it has higher peak compute throughput capabilities per core.
We note that, since performance is measured via hardware tracing, the results in Table \ref{tb:single_core_GEMM_results} include the inevitable memory stalls due to bank conflicts, thus reflecting the \textit{actual} NPU performance (when not bounded by 
DRAM BW).
Finally, very high L1 memory usage is achieved across all solutions, ranging from 94--97\%.


\begin{table*}[t]
 \centering
\caption{Evaluation of two top-ranked solutions for XDNA across various data types ($B$ \textit{column-major}).}
\setlength\tabcolsep{3.0pt}
\renewcommand{\arraystretch}{0.98}
\resizebox{0.88\textwidth}{!}{
\begin{tabular}{c|c|c|c|c|c|c||c|c}
\Xhline{2.5\arrayrulewidth}

\textbf{Precision} & \textbf{Kernel Size} & \textbf{Product} & \textbf{Thrghpt.} & \textbf{L1 Core} & \textbf{L2 Total} & \textbf{Peak} & \textbf{GEMM Size} & \textbf{Actual} \\

\textbf{In-Out} & \textbf{\textbf{$m_\text{ct}$ $\times$ $k_\text{ct}$ $\times$ $n_\text{ct}$}} & \textbf{$m_\text{ct}$ $\cdot$ $n_\text{ct}$} & \textbf{MACs/cyc} & \textbf{Mem. (KB)} & \textbf{Mem. (KB)} & \textbf{Comp. TOPS} & \textbf{$M$ $\times$ $K$ $\times$ $N$} & \textbf{NPU TOPS} \\

\hline
\hline

\multirow{2}{*}{\textbf{int8-int8}} & \textbf{112 $\times$ 112 $\times$ 112} & \textbf{12.3K} & \textbf{212.5}  & \textbf{61.3 (96\%)} & \textbf{980 (48\%)} & \textbf{6.80} & \textbf{4032 $\times$ 4032 $\times$ 4032} & \textbf{6.52}\\

 & 112 $\times$ 104 $\times$ 128 & 14.0K & 207.4 & 62.8 (98\%) & 1004 (49\%) & 6.63 & 4032 $\times$ 4160 $\times$ 4096 & 6.48\\

\hline

\multirow{2}{*}{\textbf{int8-int16}} & \textbf{96 $\times$ 112 $\times$ 96} & \textbf{9.0K} & \textbf{192.0} & \textbf{60.0 (94\%)} & \textbf{960 (47\%)} & \textbf{6.14} &  \textbf{4224 $\times$ 4032 $\times$ 4224} & \textbf{5.85}\\

 & 80 $\times$ 104 $\times$ 128 & 10.0K & 186.9 & 62.3 (97\%) & 996 (49\%) & 5.98 & 4160 $\times$ 4160 $\times$ 4096 & 5.75 \\

\hline

\multirow{2}{*}{\textbf{int8-int32}} & \textbf{80 $\times$ 88 $\times$ 96} & \textbf{7.5K} & \textbf{146.0} & \textbf{60.3 (94\%)} & \textbf{964 (47\%)} & \textbf{4.67} &  \textbf{4160 $\times$ 4224 $\times$ 4224} & \textbf{4.42}\\

 & 64 $\times$ 80 $\times$ 128 & 8.0K & 133.1 & 62.0 (97\%) & 992 (48\%) & 4.26 & 4096 $\times$ 4160 $\times$ 4096 & 4.09 \\

\hline

\multirow{2}{*}{\textbf{bf16-bf16}} & \textbf{96 $\times$ 56 $\times$ 96} & \textbf{9.0K} & \textbf{99.8}  & \textbf{60.0 (94\%)} & \textbf{960 (47\%)} & \textbf{3.19} &  \textbf{4224 $\times$ 4032 $\times$ 4224} & \textbf{3.12}\\

 & 96 $\times$ 48 $\times$ 112 & 10.5K & 97.3 & 60.0 (94\%) & 960 (47\%) & 3.11 & 4224 $\times$ 4032 $\times$ 4032 & 3.02 \\

\Xhline{2.5\arrayrulewidth}

\end{tabular}
}
\label{tb:top_ranked_solutions_GEMM_XDNA_array}


\end{table*}

\begin{table*}[t]
 \centering
\caption{Evaluation of two top-ranked solutions for XDNA2 across various data types ($B$ \textit{column-major}).}
\setlength\tabcolsep{3.0pt}
\renewcommand{\arraystretch}{0.98}
\resizebox{0.88\textwidth}{!}{
\begin{threeparttable}
\begin{tabular}{c|c|c|c|c|c|c||c|c}
\Xhline{2.5\arrayrulewidth}

\textbf{Precision} & \textbf{Kernel Size} & \textbf{Product} & \textbf{Thrghpt.} & \textbf{L1 Core} & \textbf{L2 Total} & \textbf{Peak} & \textbf{GEMM Size} & \textbf{Actual} \\

\textbf{In-Out} & \textbf{\textbf{$m_\text{ct}$ $\times$ $k_\text{ct}$ $\times$ $n_\text{ct}$}} & \textbf{$m_\text{ct}$ $\cdot$ $n_\text{ct}$} & \textbf{MACs/cyc} & \textbf{Mem. (KB)} & \textbf{Mem. (KB)} & \textbf{Comp. TOPS} & \textbf{$M$ $\times$ $K$ $\times$ $N$} & \textbf{NPU TOPS} \\

\hline
\hline

\multirow{2}{*}{\textbf{int8-int8}} & \textbf{144 $\times$ 72 $\times$ 144} & \textbf{20.3K} & \textbf{343.0} & \textbf{60.8 (95\%)} & \textbf{2106 (51\%)} & \textbf{39.52} & \textbf{4032 $\times$ 4320 $\times$ 4608} & \textbf{37.35}\\

 & 160 $\times$ 64 $\times$ 144 & 22.5K & 322.6 & 60.5 (95\%) & 2064 (50\%) & 37.16 & 4480 $\times$ 4224 $\times$ 4608 & 36.13\\
\hline

\multirow{2}{*}{\textbf{int8-int16}} & \textbf{128 $\times$ 72 $\times$ 112} & \textbf{14.0K} & \textbf{307.2} & \textbf{61.8 (97\%)} & \textbf{2084 (51\%)} & \textbf{35.39} &  \textbf{4096 $\times$ 4320 $\times$ 4480} & \textbf{30.77}\\

 & 160 $\times$ 64 $\times$ 96 & 15.0K & 271.4 & 62.0 (97\%) & 2016 (49\%) & 31.26 & 4480 $\times$ 4224 $\times$ 4608 & 29.59 \\
\hline

\multirow{2}{*}{\textbf{int8-int32}} & \textbf{96 $\times$ 64 $\times$ 96} & \textbf{9.0K} & \textbf{256.0} & \textbf{60.0 (94\%)} & \textbf{2016 (49\%)} & \textbf{29.49} &  \textbf{4224 $\times$ 4224 $\times$ 4608} & \textbf{24.74}\\

 & 128 $\times$ 56 $\times$ 80 & 10.0K & 209.9 & 62.3 (97\%) & 2036 (50\%) & 24.18 & 4096 $\times$ 4032 $\times$ 4480 & 21.67 \\

\hline

\multirow{2}{*}{\textbf{bf16-bf16}} & \textbf{112 $\times$ 48 $\times$ 96} & \textbf{10.5K} & \textbf{137.2} & \textbf{60.0 (94\%)} & \textbf{2496 (61\%)} & \textbf{15.81} &  \textbf{4032 $\times$ 4224 $\times$ 4608} & \textbf{14.52}\\

 & 160 $\times$ 40 $\times$ 80 & 12.5K & 124.1 & 62.5 (98\%) & 2400 (59\%) & 14.30 & 4480 $\times$ 4160 $\times$ 4480 & 13.67 \\

\Xhline{2.5\arrayrulewidth}

\end{tabular}


\end{threeparttable}
}
\label{tb:top_ranked_solutions_GEMM_XDNA2_array}


\end{table*}

\subsection{GEMM Performance on NPU Array}
\label{subsec:npu_array_GEMM_performance}

In this section, we present GEMM performance on the entire NPU array.
Performance is evaluated using \textit{wall-clock} time, thereby capturing the \textit{actual} performance observed by the users (includes OS overheads, NPU dispatch time, etc.)  \cite{wall_clock_time_mlir_aie}.
All reported results represent the average of 100 runs.

\subsubsection{\textbf{Optimal Balanced GEMM Kernel}} 
\label{subsubsec:optimal_balanced_GEMM_kernel}


As shown in the previous section, GEMM kernels can attain very high throughput.
However, when using the optimum compute kernel sizes of Table \ref{tb:single_core_GEMM_results}, we observe that GEMM performance on the NPU array remains low.
For instance, for int8-int16 precision, we obtain only 17.86 TOPS on XDNA2 at $\sim$4K square GEMM size. 
However, the peak compute capability of this kernel on the XDNA2 array is 48.36 TOPS, when calculated at maximum operating frequency (identified through XDNA driver commands \cite{xdna_driver_github_repo}, \emph{i.e.,} 1 GHz and 1.8 GHz for XDNA and XDNA2, respectively).
This implies that GEMM is memory bound at this specific kernel size, due to low values of $m_\text{ct}$ and $n_\text{ct}$ (inverse relationship of compute and memory). 
Therefore, we leverage the optimization methodology described in Sec. \ref{subsubsec:system_level_optimization} in order to identify the optimal \textit{balanced} kernel (\emph{i.e.,} where compute and memory become balanced).
First, we use micro-benchmarking to estimate the \textit{effective} DRAM BW that is available to the NPU when running GEMM workloads.
For micro-benchmarking, we imitate GEMM transfers from DRAM to NPU array and vice-versa, observing $\sim$15 GB/s and $\sim$50 GB/s for XDNA and XDNA2, respectively.
Afterwards, we exploit these values to select starting points for the iterative optimization procedure of Sec. \ref{subsubsec:system_level_optimization}.


Tables \ref{tb:top_ranked_solutions_GEMM_XDNA_array} and \ref{tb:top_ranked_solutions_GEMM_XDNA2_array} present the two top-ranked solutions, for XDNA and XDNA2, respectively.
First, notice that kernel sizes have lower $k_\text{ct}$ and higher $m_\text{ct}$, $n_\text{ct}$ compared to kernels in Table \ref{tb:single_core_GEMM_results}.
This decreases their compute throughput; for example, 96$\times$112$\times$96 for int8-int16 (Table \ref{tb:top_ranked_solutions_GEMM_XDNA_array}) achieves 192.0 MACs/cycle, compared to 217.6 MACs/cycle of 64$\times$216$\times$64 (Table \ref{tb:single_core_GEMM_results}).
However, GEMM performance on the NPU array increases, since compute and memory become balanced.
For example, when using the 96$\times$112$\times$96 kernel, GEMM performance on XDNA for $\sim$4K GEMM size is 5.85 TOPS (Table \ref{tb:top_ranked_solutions_GEMM_XDNA_array})  compared to 4.03 TOPS of 64$\times$216$\times$64 (not shown in Table \ref{tb:single_core_GEMM_results}). 



Second, observe that when further increasing the product of $m_\text{ct} \cdot n_\text{ct}$ (thus decreasing $k_\text{ct}$ to fit in L1), GEMM performance on NPU array drops.
For instance, for int8-int8, kernel 160$\times$64$\times$144 has higher $m_\text{ct} \cdot n_\text{ct}$ product compared to 144$\times$72$\times$144 (22.5K \emph{vs.} 20.3K), but performance is lower: 36.13 \emph{vs.} 37.35 TOPS (Table \ref{tb:top_ranked_solutions_GEMM_XDNA2_array}).
This is because at this point GEMM has become compute bound, but compute has also become lower (throughput drops to 322.6 from 343.0 MACs/cycle).
The peak compute throughput on the entire XDNA2 array when attaining single-core throughput of 322.6 MACs/cycle is 37.16 TOPS.
However, the actual GEMM performance is 36.13 TOPS for $\sim$4K GEMM size, as shown in Table \ref{tb:top_ranked_solutions_GEMM_XDNA2_array}.
This $\sim$3\% difference is mainly attributed to DMA transfers of $C$ tiles (single output buffer), occurring at the end of each complete reduction across $K$ (every $K$/$k_\text{ct}$ tiles).
Another smaller contribution  comes from the vectorized zeroing kernel executing every $K$/$k_\text{ct}$ tiles (Sec. \ref{subsec:GEMM_mapping_strategy}), which is typically $<$10\% of GEMM kernel time (thus, $<$0.15\% for the $\sim$4K GEMM of this particular example).
Both output DMA transfers and zeroing kernel cycles have been verified using NPU tracing.
Note that as $K$ becomes higher, these effects are further amortized.
Other contributions include the DRAM transfer time for initial $A$ and $B$ tiles, along with final output $C$ tiles, as well as overheads due to \textit{wall-clock} time measurements (\emph{e.g.,} NPU dispatch time) \cite{wall_clock_time_mlir_aie}.
In summary, the bolded solutions across all precisions in Tables \ref{tb:top_ranked_solutions_GEMM_XDNA_array} and \ref{tb:top_ranked_solutions_GEMM_XDNA2_array} represent the optimal \textit{balanced} point between compute and memory, where GEMM performance is maximized.

We note that results in Tables \ref{tb:top_ranked_solutions_GEMM_XDNA_array}, \ref{tb:top_ranked_solutions_GEMM_XDNA2_array} correspond to matrix $B$ in \textit{column-major} (results for $B$ in \textit{row-major} are presented in Sec. \ref{subsubsec:GEMM_sweeps}). 
Across all results, the contiguous parameter $k_\text{mt}$, which is utilized to increase the effective DRAM BW  (and hence GEMM performance), is configured as shown below (Sec. \ref{subsubsec:contiguous_kmt_parameter}).
Finally, the optimization procedure requires less than 30 minutes to identify the optimum solution for each data type (evaluated on the corresponding mini PCs).
The overall execution time is dominated by the compilation time of \emph{xchesscc} and IRON (up to 5 minutes per iteration), while the exhaustive search takes less than 1 s in all cases. 



\subsubsection{\textbf{Contiguous $k_\text{mt}$ Parameter}} 
\label{subsubsec:contiguous_kmt_parameter}

\begin{figure}[t]
\vspace{-0.40cm}
\centering
\subfloat[XDNA]
{\includegraphics[width=0.49\linewidth]{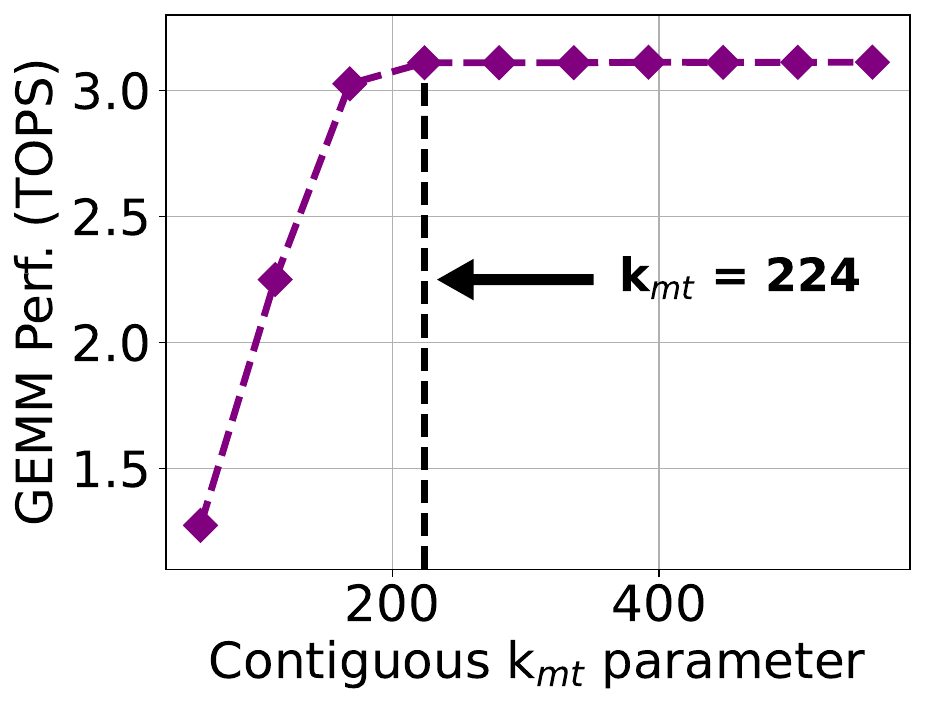}
\label{fig:contiguous_bf16_bf16_XDNA}}
\hfill
\subfloat[XDNA2]{\includegraphics[width=0.485\linewidth]{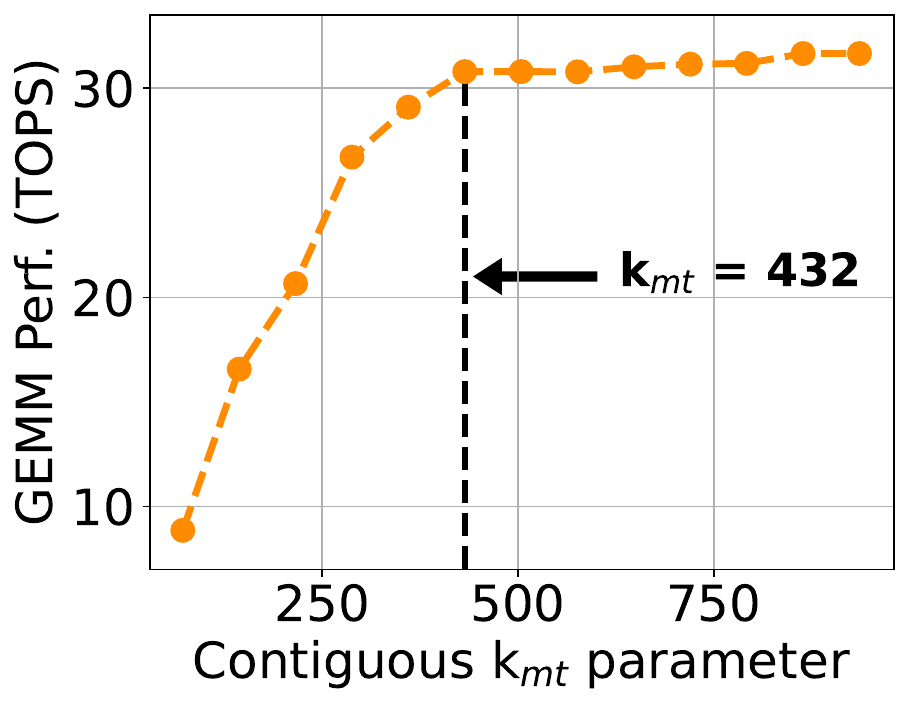}
\label{fig:contiguous_int8_int16_XDNA2}} 

\vspace{-0.20cm}

\caption{GEMM performance while varying parameter $k_\text{mt}$ for bf16-bf16 96$\times$56$\times$96 (a) and int8-int16 128$\times$72$\times$112 (b).} 
\label{fig:GEMM_performance_vs_contiguous_param_k_mt}
\vspace{-0.40cm}
\end{figure}

In this section, we present the impact of the contiguous $k_\text{mt}$ parameter on GEMM performance.
Fig. \ref{fig:GEMM_performance_vs_contiguous_param_k_mt} depicts the GEMM performance when varying the parameter $k_\text{mt}$ and utilizing the optimal \textit{balanced} kernels for two arbitrary data types ($\sim$4K GEMM size and $B$ in \textit{column-major}).
For instance, for XDNA, when $k_\text{mt}$ is equal to $k_\text{ct}$ ($=$ 56 in Fig. \ref{fig:contiguous_bf16_bf16_XDNA}), GEMM performance is very low (\emph{i.e.,} 1.27 TOPS).
As $k_\text{mt}$ increases (in multiples of $k_\text{ct}$) performance improves.
This is because it enables higher \textit{effective} DRAM BW, by traversing more contiguous elements for both $A$ (\textit{row-major}) and $B$ (\textit{column-major}) matrices. 
However, at some point, this enhancement becomes saturated (\emph{i.e.,} higher $k_\text{mt}$ provides marginal improvement).
Therefore, we \textit{empirically} select the smaller value where GEMM performance becomes saturated for each data type (\emph{i.e.,} $k_\text{mt}$=224 in Fig. \ref{fig:contiguous_bf16_bf16_XDNA}).
Note that selecting the smallest value while also maintaining high GEMM performance is essential, since it reduces any potential zero padding needed to align the final GEMM dimensions with the \textit{native} GEMM size (Sec. \ref{subsec:GEMM_mapping_strategy}).
For example, for the bf16-bf16 case, the \textit{native} GEMM size operating natively on the entire 4$\times$4 XDNA array is 384$\times$224$\times$384.
We note here that this results in reduced L2 memory utilization, ranging from 47--61\% across all data types (Tables \ref{tb:top_ranked_solutions_GEMM_XDNA_array} \& \ref{tb:top_ranked_solutions_GEMM_XDNA2_array}).
While maximized L2 memory usage can be attained for higher $k_\text{mt}$ values (\emph{e.g.,} 96\% for $k_\text{mt}$=560 in Fig. \ref{fig:contiguous_bf16_bf16_XDNA}), this only leads to marginal performance improvement in GEMM ($<$ 1\%).

Similarly, for XDNA2, for int8-int16, we set $k_\text{mt}$=432 (Fig. \ref{fig:contiguous_int8_int16_XDNA2}).
In this case, the \textit{native} GEMM size on the XDNA2 array becomes 512$\times$432$\times$896.
We note here that the three latest points of Fig. \ref{fig:contiguous_int8_int16_XDNA2} are enabled via the neighboring memory sharing in MemTiles, as a direct result of the GEMM mapping strategy on XDNA2 (Sec. \ref{subsubsec:MemTiles_design}).
For all other data types, we set the parameter $k_\text{mt}$ in a similar fashion. 
Specifically, for XDNA, we set $k_\text{mt}$ equal to 448 for int8-int8 and int8-int16, while for int8-int32 we set it to 352.
On XDNA2, we use 432 for int8-int8, and 384 for both int8-int32 and bf16-bf16.

It is important to mention here that the non-optimized GEMM example in \cite{GEMM_example_mlir_aie}, cannot support sufficient contiguous elements.
For a fair comparison, we utilize our optimized \textit{balanced} kernels and modify their implementation to support single output $C$ buffers (allowing it to fit in L1).
For instance, for the data types shown in Fig. \ref{fig:GEMM_performance_vs_contiguous_param_k_mt}, we attain 2.4$\times$ and 3.6$\times$ higher performance for XDNA and XDNA2, respectively.
These results highlight the importance of accessing sufficient contiguous elements in GEMM performance.

\subsubsection{\textbf{GEMM Performance Sweeps}} 
\label{subsubsec:GEMM_sweeps}



\begin{figure*}[ht]
\vspace{-0.50cm}
\centering
\subfloat[int8-int8]
{\includegraphics[width=0.28\textwidth]{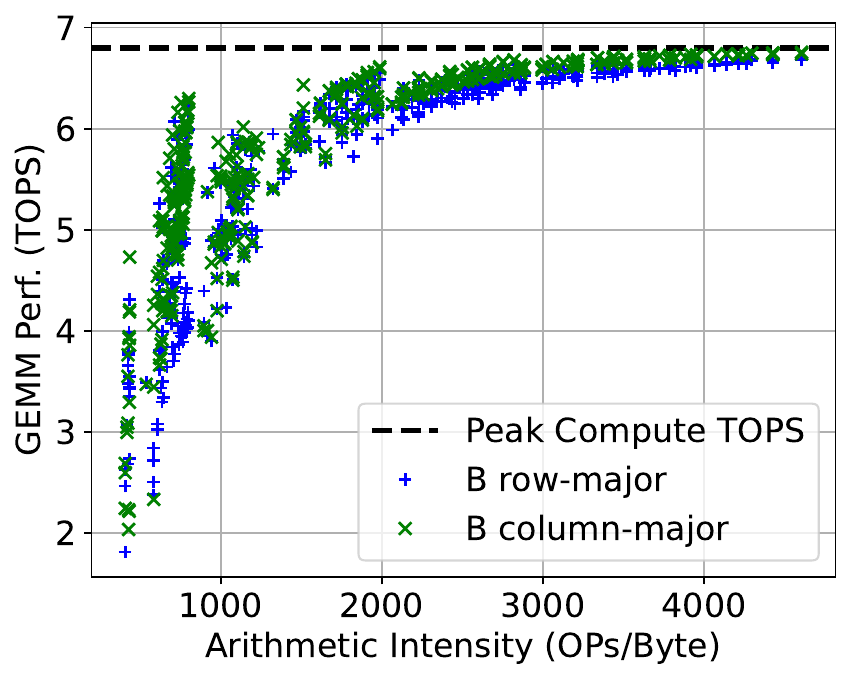}
\label{fig:GEMM_sweep_int8_int8_phoenix}}
\hspace{8mm}
\subfloat[int8-int16]{\includegraphics[width=0.28\textwidth]{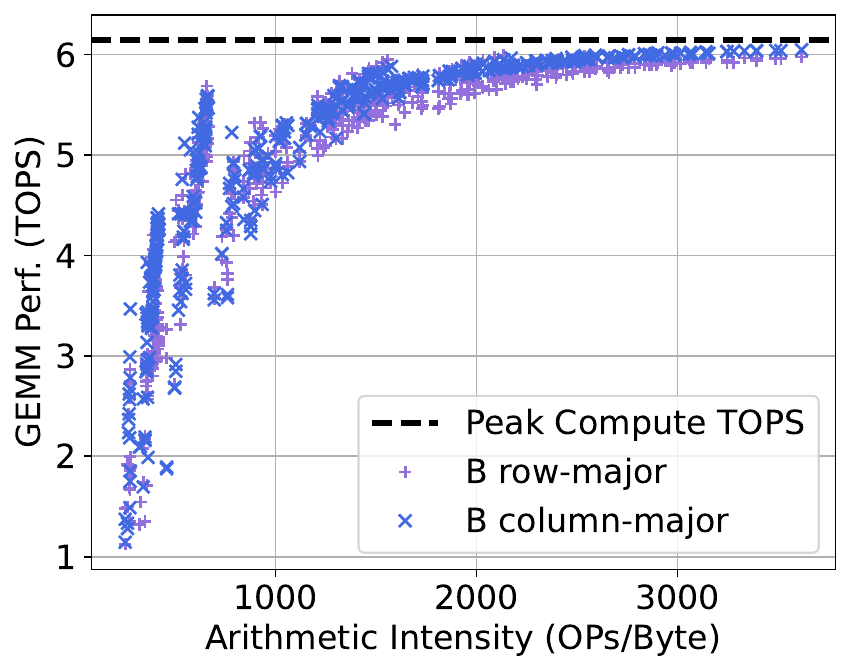}
\label{fig:GEMM_sweep_int8_int16_phoenix}} 
\hspace{8mm}
\subfloat[bf16-bf16]{\includegraphics[width=0.29\textwidth]{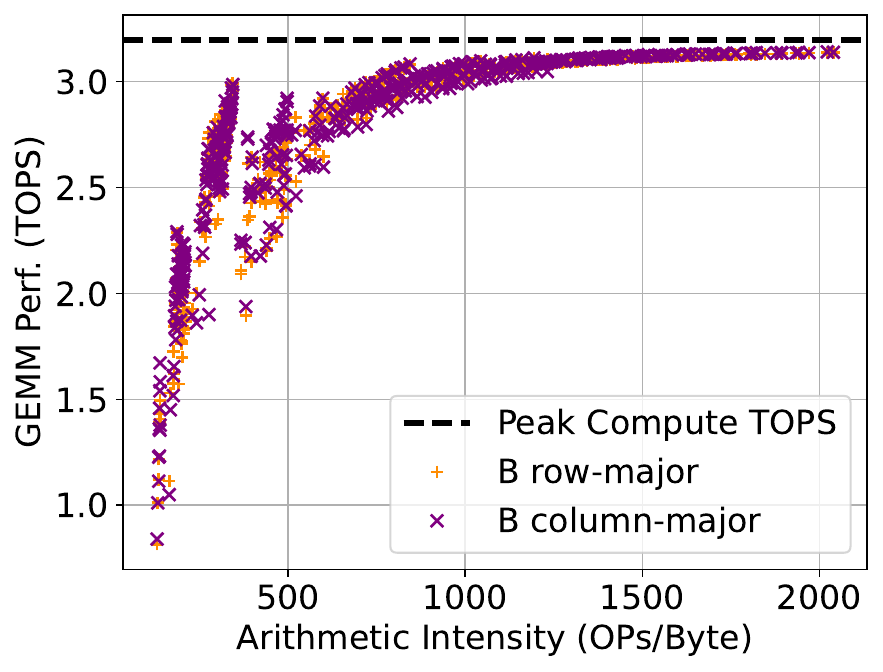}
\label{fig:GEMM_sweep_bf16_bf16_phoenix}} 

\vspace{-0.30cm}

\caption{Roofline GEMM performance sweeps for various matrix sizes on XDNA.} 
\label{fig:GEMM_sweeps_phoenix}
\end{figure*}

\begin{figure*}[ht]
\vspace{-0.50cm}
\centering
\subfloat[int8-int8]
{\includegraphics[width=0.28\textwidth]{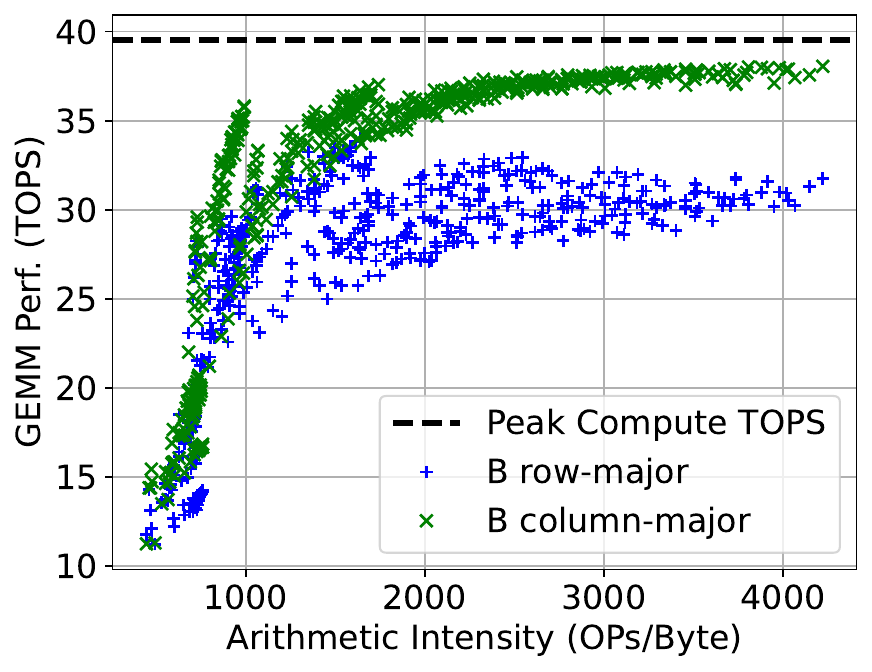}
\label{fig:GEMM_sweep_int8_int8_kracken}}
\hspace{8mm}
\subfloat[int8-int16]{\includegraphics[width=0.28\textwidth]{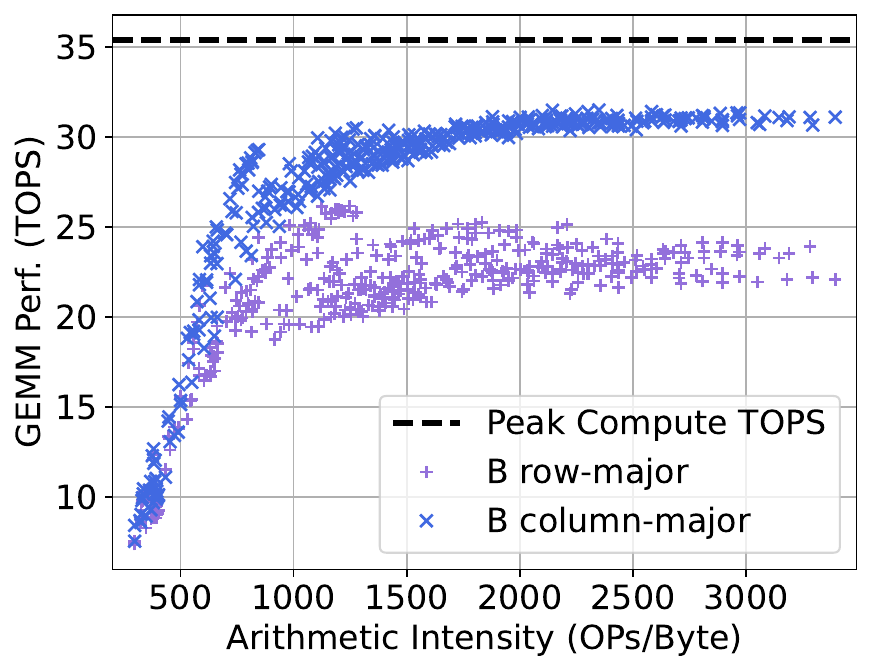}
\label{fig:GEMM_sweep_int8_int16_kracken}} 
\hspace{8mm}
\subfloat[bf16-bf16]{\includegraphics[width=0.28\textwidth]{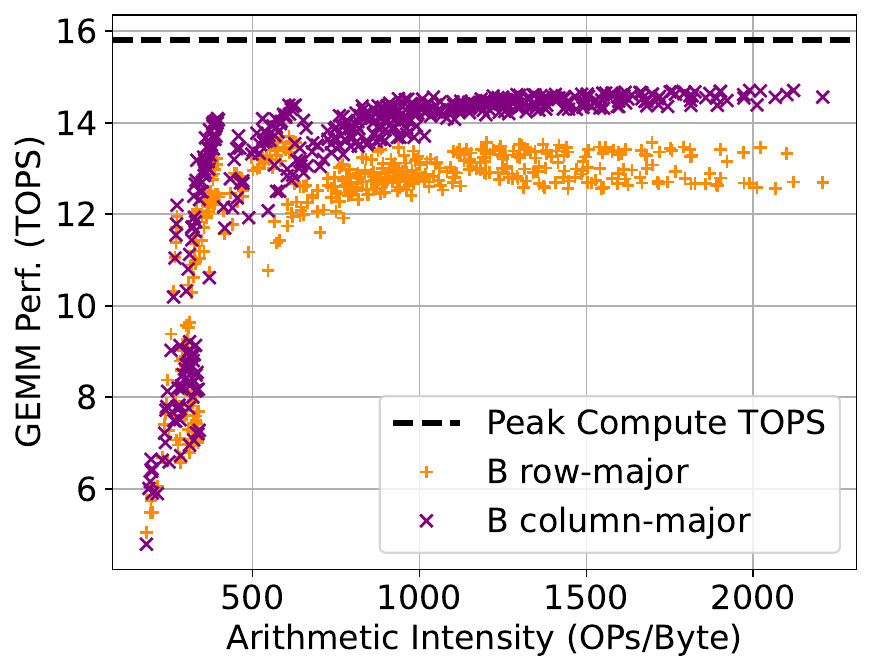}
\label{fig:GEMM_sweep_bf16_bf16_kracken}} 

\vspace{-0.30cm}

\caption{Roofline GEMM performance sweeps for various matrix sizes on XDNA2.} 
\label{fig:GEMM_sweeps_kracken}
\vspace{-0.20cm}
\end{figure*}

In Fig. \ref{fig:GEMM_sweeps_phoenix} and \ref{fig:GEMM_sweeps_kracken}, we show roofline GEMM performance sweeps (in linear scale).
Each point represents a matrix size that is a multiple of the \textit{native} GEMM size (using the optimal \textit{balanced} kernel).
We select more than 400 points for each case (separately for $B$ in \textit{column-} and \textit{row-major}), up to 8K-sized matrices, without favoring any particular $M$, $K$, $N$ dimension.
First, notice that when arithmetic intensity (ARI) is low (\emph{i.e.,} small matrix sizes), performance is limited.
In this case, GEMM is severely memory bound.
However, as ARI increases, GEMM performance improves, and becomes more stabilized after a specific value. 
In all cases, storing matrix $B$ in \textit{column-major} provides, on average, higher performance compared to \textit{row-major}.
This is because of accessing sufficient contiguous data for both $A$ and $B$ matrices (determined by $k_\text{mt}$ parameter).
However, for $B$ in \textit{row-major}, the contiguous access is limited to the $n_\text{ct}$ parameter, while only for $A$ (\textit{row-major}), $k_\text{mt}$ contiguous data are traversed.
To this end, for XDNA, we observe, on average, 4.8\%, 4.4\%, and 0.57\% higher performance, for int8-int8, int8-int16, and bf16-bf16, respectively.

Moreover, we notice that for XDNA2 the difference between \textit{column-} and \textit{row-major} is higher.
In particular, we observe, on average, 19.1\%, 25.2\%, and 8.7\% higher performance, for int8-int8, int8-int16, and bf16-bf16, respectively.
This difference between XDNA and XDNA2 is presumably attributed to complex interaction between the NPU NoC, the SoC-level fabric and DRAM \cite{NPU_arch_MICRO_2024, NPU_HOT_CHIPS_2023}, which affects the \textit{effective} DRAM BW that NPU perceives (although both mini PC devices are equipped with the same DRAM).
Also, we note that for both XDNA and XDNA2, the difference between \textit{column-} and \textit{row-major} for bf16 is lower compared to int8.
This is because of accessing a larger number of bytes for bf16 across the $n_\text{ct}$ dimension (when $B$ is in \textit{row-major}), which increases GEMM performance in this case, thereby lowering their difference.


For XDNA, we note that performance gets stabilized  after a specific ARI value for $B$ in both \textit{row-} and \textit{column-major} formats (almost resembling a line in Fig. \ref{fig:GEMM_sweeps_phoenix}).
Moreover, for XDNA2, we observe that for $B$ in \textit{column-major}, GEMM performance also resembles a stable line.
However, for $B$ in \textit{row-major} it displays a more scattered distribution (Fig. \ref{fig:GEMM_sweeps_kracken}).
This is because XDNA2 has higher reliance on the  \textit{effective} DRAM BW (due to attaining significantly higher absolute TOPS values), which is substantially increased and stabilized when accessing sufficient contiguous data across both $A$ and $B$ matrices.
For example, for int8-int16 on XDNA2, we measure a variability of only 5\% for $B$ in \textit{column-major}, while for \textit{row-major} the variability is 19\% (ARI > 1600).
Finally, across all points in GEMM sweeps, XDNA attains up to 6.76 (int8-int8), 6.05 (int8-int16), 4.57 (int8-int32), and 3.14 (bf16-bf16) TOPS.
Similarly, XDNA2 achieves up to 38.05 (int8-int8), 31.52 (int8-int16), 25.31 (int8-int32), and 14.71 (bf16-bf16) TOPS (int8-int32 sweep omitted for brevity).




\subsection{Insights \& Discussion}
\label{subsec:insights_and_discussion}

\subsubsection{\textbf{Performance Across Multiple GEMM Sizes in DL Workloads}} 
\label{subsubsec:GEMM_perf_multiple_sizes}


Modern DL workloads perform GEMMs with a wide range of  sizes across their layers. 
Our employed output stationary mapping allows \textit{arbitrary} GEMM dimensions to be supported, by applying zero-padding to align with the \textit{native} GEMM size (Sec. \ref{subsec:GEMM_mapping_strategy}).
Zero-padding can be applied efficiently by utilizing the NPU's architectural support for on-the-fly zero-padding in MemTile channels \cite{AMD_AIE_ML_architecture_manual}; leveraging this feature is left for future work.

Switching between different GEMM sizes can incur critical performance inefficiencies.
To this end, one approach could be to reconfigure the NPU array with a dedicated GEMM design for each size.
When quantifying the reconfiguration latency of the entire GEMM design, we measure a delay of 3.4 ms and 4.9 ms on XDNA and XDNA2, respectively.
However, this reconfiguration latency is comparable to the GEMM execution time (\emph{e.g.,} a $\sim$4K square GEMM for int8-int16 on XDNA2 takes 5.2 ms).
This underscores that reconfiguring the entire design can impose substantial overheads.

When retaining the \textit{same} GEMM design on the NPU, only two parameters require reconfiguration across different problem sizes ($M, K, N$): (i) the total number of output tiles $M\cdot N/(m_\text{ct} \cdot n_\text{ct})$, and (ii) the number of tiles across the reduction dimension $K/k_\text{ct}$ \cite{roesti2025unlocking}.
According to our measurements, this negligible reconfiguration does not incur any noticeable performance overhead at the system level.
Hence, identifying the optimal parameters (\emph{i.e.,} $m_\text{ct}$, $k_\text{ct}$, $n_\text{ct}$, $k_\text{mt}$, Sec. \ref{sec:GEMM_Design}), and reusing them across different GEMM sizes is essential for high-performance DL deployment.
Finally, we note that the system-level GEMM results presented are specific to the mini PCs used in this evaluation.
However, our optimization methodology is generalizable to any NPU device in the current two generations.


\subsubsection{\textbf{Single Output Buffer}} 
\label{subsubsec:single_output_buffer}

Due to output stationary mapping, we retain the output $C$ tiles as single buffers and apply double-buffering only for inputs $A$ and $B$ (Sec. \ref{subsec:GEMM_mapping_strategy}).
This is crucial in maximizing performance in the \textit{general} GEMM case, because it enables significantly higher flexibility in single-core parameter optimization.
In particular, it allows increased values for $m_\text{ct}$ $\times$ $k_\text{ct}$ $\times$ $n_\text{ct}$ parameters, compared to utilizing double-buffering (constrained by L1 memory size).
This results in identifying a \textit{balanced} kernel that has higher performance, thereby enabling higher GEMM performance at the system level.
For example, when using the optimization methodology described in Sec. \ref{subsubsec:system_level_optimization} and apply double-buffering for $C$, we identify the 112$\times$48$\times$96 size as the optimal \textit{balanced} kernel for int8-int16 data type on XDNA2.
For $\sim$4K square GEMM, this kernel provides 26.1 TOPS.
However, the 128$\times$72$\times$112 kernel of Table \ref{tb:top_ranked_solutions_GEMM_XDNA2_array} (single $C$ buffer), provides 30.77 TOPS, representing an 18\%  performance improvement.
Similarly, on XDNA, double buffer on $C$ provides 2.76 TOPS (80$\times$40$\times$96 kernel), while single buffer offers 3.12 TOPS (96$\times$56$\times$96 kernel on Table \ref{tb:top_ranked_solutions_GEMM_XDNA_array}, 13\% higher).
The transfer of the output $C$ tiles in the single  buffer case gets amortized as the reduction $K$ dimension becomes sufficiently high (typically $<$5\% degradation in GEMM performance when $K$/$k_\text{ct}$$>$20).





\subsubsection{\textbf{NPU--DRAM Data Movement \& BD Reconfiguration}} 
\label{subsubsec:BD_reconfiguration}

The procedure delineated in Sec. \ref{subsec:npu_main_memory_data_movement} enables efficient overlapping of NPU--DRAM data movement with BD reconfiguration.
To quantify its impact on performance, we modify our design to synchronize and reconfigure BDs sequentially (without overlap).
In this case, for int8-int16 on XDNA2, we notice only 22.21 TOPS at $\sim$4K square GEMM, while the overlapped design of Table \ref{tb:top_ranked_solutions_GEMM_XDNA2_array} exhibits 30.77 TOPs (28\% decrease for non-overlapped design).
Similarly, for XDNA, for int8-int16 (Table \ref{tb:top_ranked_solutions_GEMM_XDNA_array}), we observe a 27\% degradation in performance.
This highlights the critical role of overlapping DMA transfers with BD reconfiguration in attaining high GEMM performance.




\subsubsection{\textbf{Future Research}} 
\label{subsubsec:Future Work}

XDNA2 incorporates hardware support for bfp16 precision, where multiple numbers share one common exponent.
This incurs additional challenges for data layout transformations exploiting the multi-dimensional DMA addressing features of the NPUs (Sec. \ref{subsec:DMA_transformations}).
However, this is beyond the scope of this paper and will therefore be addressed in future work.
Furthermore, our proposed optimization methodology can be also be exploited for special cases of GEMM such as general matrix-vector multiplication (GEMV), which we also leave as future work.




\section{Conclusion}
\label{sec:Conclusion}


In this work, we propose a novel optimization methodology to maximize GEMM performance on Ryzen AI NPUs.
We observe the \textit{inverse} relationship between compute and off-chip memory and determine the optimal balanced point, where performance is maximized.
To identify this optimal performance point, we exploit analytical modeling and hardware profiling techniques.
Our methodology attains state-of-the-art GEMM performance and is generalizable across the current AMD Ryzen AI NPU generations.




\bibliographystyle{ACM-Reference-Format}
\bibliography{bibliography}

\end{document}